\RequirePackage{fixltx2e} 
\documentclass[aps,prd,superscriptaddress,showpacs,nofootinbib,floatfix,twocolumn]{revtex4}

\usepackage{amsmath}
\usepackage{amssymb}
\usepackage{graphicx, subfigure}
\usepackage{array,longtable}
\usepackage[english]{babel}
\usepackage{multirow}
\usepackage{xspace}
\usepackage{booktabs}
\usepackage[sc]{mathpazo}
\usepackage{mathtools}

\begin{document}

\title{Inferences on Mass Composition and Tests of Hadronic
Interactions from 0.3 to 100 EeV using the water-Cherenkov Detectors of
the Pierre Auger Observatory}


\author{A.~Aab}
\affiliation{IMAPP, Radboud University Nijmegen, Nijmegen, The Netherlands}

\author{P.~Abreu}
\affiliation{Laborat\'orio de Instrumenta\c{c}\~ao e F\'\i{}sica Experimental de Part\'\i{}culas -- LIP and Instituto Superior T\'ecnico -- IST, Universidade de Lisboa -- UL, Lisboa, Portugal}

\author{M.~Aglietta}
\affiliation{Osservatorio Astrofisico di Torino (INAF), Torino, Italy}
\affiliation{INFN, Sezione di Torino, Torino, Italy}

\author{I.~Al Samarai}
\affiliation{Laboratoire de Physique Nucl\'eaire et de Hautes Energies (LPNHE), Universit\'es Paris 6 et Paris 7, CNRS-IN2P3, Paris, France}

\author{I.F.M.~Albuquerque}
\affiliation{Universidade de S\~ao Paulo, Instituto de F\'\i{}sica, S\~ao Paulo, SP, Brazil}

\author{I.~Allekotte}
\affiliation{Centro At\'omico Bariloche and Instituto Balseiro (CNEA-UNCuyo-CONICET), San Carlos de Bariloche, Argentina}

\author{A.~Almela}
\affiliation{Instituto de Tecnolog\'\i{}as en Detecci\'on y Astropart\'\i{}culas (CNEA, CONICET, UNSAM), Buenos Aires, Argentina}
\affiliation{Universidad Tecnol\'ogica Nacional -- Facultad Regional Buenos Aires, Buenos Aires, Argentina}

\author{J.~Alvarez Castillo}
\affiliation{Universidad Nacional Aut\'onoma de M\'exico, M\'exico, D.F., M\'exico}

\author{J.~Alvarez-Mu\~niz}
\affiliation{Universidad de Santiago de Compostela, Santiago de Compostela, Spain}

\author{G.A.~Anastasi}
\affiliation{Gran Sasso Science Institute (INFN), L'Aquila, Italy}
\affiliation{INFN Laboratori Nazionali del Gran Sasso, Assergi (L'Aquila), Italy}

% ...

\collaboration{The Pierre Auger Collaboration}
\email{auger_spokespersons@fnal.gov}
\homepage{http://www.auger.org}
\noaffiliation

\begin{abstract}

We present a new method for probing the hadronic interaction models at
ultra-high energy and extracting details about mass composition. This is
done using the time profiles of the signals recorded with the
water-Cherenkov detectors of the Pierre Auger Observatory. The profiles
arise from a mix of the muon and electromagnetic components of
air-showers. Using the risetimes of the recorded signals we define a new
parameter, which we use to compare our observations with predictions
from simulations. We find, firstly, inconsistencies between our data and
predictions over a greater energy range and with substantially more
events than in previous studies. Secondly, by calibrating the new
parameter with fluorescence measurements from observations made at the
Auger Observatory, we can infer the depth of shower maximum
\textit{X}\textsubscript{max} for a sample of over 81,000 events extending
from 0.3 EeV to over 100 EeV. Above 30 EeV, the sample is nearly
fourteen times larger than currently available from fluorescence
measurements and extending the covered energy range by half a decade. 
The energy dependence of $\langle$\textit{X}\textsubscript{max}$\rangle$
is compared to simulations and interpreted in terms of the mean of the
logarithmic mass. We find good agreement with previous work and extend
the measurement of the mean depth of shower maximum to greater energies
than before, reducing significantly the statistical uncertainty associated with 
the inferences about mass composition.
\end{abstract}

\pacs{96.50.S-, 96.50.sb, 96.50.sd, 98.70.Sa}

\maketitle

\section{Introduction}

Our understanding of the properties of the highest-energy cosmic rays
has grown enormously over the last 12 years with the advent of data from
the Pierre Auger Observatory and the Telescope Array. These devices have
been used to study the energy spectrum, the mass composition and the
distribution of arrival directions of cosmic rays from 0.3 EeV to
energies beyond 10 EeV. While the features of the energy spectrum and of
the arrival directions have been well-characterised up to
$\sim$100 EeV, the situation with regard to the mass spectrum
is less satisfactory because of the reliance on models of the hadronic
physics. This state of affairs arises for two reasons. Firstly, the
method that provides the best resolution, and therefore is the most
potent for measuring a mass-sensitive feature of extensive air-showers,
is the fluorescence technique. It has been exploited on an
event-by-event basis to determine the depth of shower maximum, i.e. the
depth in the atmosphere at which the energy deposition in the shower is
greatest, but, as observations are restricted to clear moonless-nights,
the number of events is limited. For example, in the Auger data so far
reported (up to 31 December 2012) there are 227 events above 16 EeV \cite{r1}.
For the same energy range, the event number from the Telescope Array is
smaller, 25 \cite{r2}. Secondly, to interpret the data sets from
the water-Cherenkov detectors and the fluorescence telescopes, one must
use the predictions of features of hadronic interactions, such as the
cross-sections for proton and pion interactions, the multiplicity and
the inelasticity, at centre-of-mass energies up to $\sim$300
TeV, well-beyond what is accessible at the LHC \cite{r3}.

To overcome the limitations imposed by the relatively small number of
events accumulated with the fluorescence technique at the highest
energies, use can be made of data recorded with the water-Cherenkov
detectors of the Observatory which are operational nearly 100\% of the
time and thus yield substantially more events at a given energy. In this
paper, we describe a new method for extracting information about the
development of air showers from the time profiles of the signals from
the water-Cherenkov detectors.

Our method allows a comparison of the data with predictions from models
of parameters inferred directly from these detectors in which a signal
from a mix of the muon and electromagnetic components of air-showers is
available. This approach follows the line opened in four recent studies.
From comparisons of Auger observations with hadronic models it is argued
that the latter are inadequate to describe the various measurements
\cite{r4,r5,r6,r7}. 
As in the earlier work\textbf{,} we find that there are inconsistencies
between the models and the data: this is established over a greater
energy range and with more events than before.

The method also enables us to infer the depth of shower maximum (a
dominantly electromagnetic measurement) by calibrating the new parameter
with measurements from the fluorescence telescopes. We have determined
\textit{X}\textsubscript{max} for about three times more events than
available from these telescopes alone over the range from
$\sim$0.3 EeV to beyond 70 EeV: specifically for the two
surface detector configurations, the 750 and 1500 m arrays
\cite{r8}, there are 27553 and 54022 events recorded respectively
for which estimates of \textit{X}\textsubscript{max} have been possible.
Of those, 49 events are in the range beyond 70 EeV, and 1586 above 20
EeV.

The structure of the paper is as follows. In section \ref{s2} features of the
Pierre Auger Observatory are briefly outlined. The measurement of the
risetime of signals from the water-Cherenkov detectors is described in
section \ref{s3} and the new parameter for studying the depth of shower maximum
is introduced in section \ref{s4}. A comparison of the new parameter with
predictions from hadronic models is discussed in section \ref{s5}. Section \ref{s6}
presents the results on the measurement of average
\textit{X}\textsubscript{max} from 0.3 EeV to beyond 70 EeV. A summary of
the conclusions is given in section \ref{s7}.

\section{The Pierre Auger Observatory}
\label{s2}

The Pierre Auger Observatory is located near the city of Malarg\"ue in the
Province of Mendoza, Argentina. It is a hybrid system, being a
combination of a large array of surface detectors and a set of
fluorescence detectors, used to study cosmic rays with energies above
0.1 EeV. Full details of the instrumentation and the methods used for
event reconstruction are given in \cite{r8}.

The work presented here is based on data gathered from 1 Jan 2004 to 31
Dec 2014 from the surface-detector array (SD), which covers an area of
over 3000 km\textsuperscript{2}. The array contains 1660 water-Cherenkov
detectors, 1600 of which are deployed on a hexagonal grid with 1500 m
spacing with the remainder on a lattice of 750 m covering 23.5
km\textsuperscript{2}. We refer to these configurations as the 1500 m
and 750 m arrays. The water-Cherenkov detectors are used to sample the
electromagnetic and muonic components of extensive air-showers. Each
detector contains 12 tonnes of ultra-pure water in a cylindrical
container, lined with Tyvec, 1.2 m deep and of 10 m\textsuperscript{2}
area. The water is viewed by three 9''-photomultiplier tubes (PMTs).
Signals from the anode and an amplified signal from the last dynode of
each PMT are digitised using 40 MHz 10-bit Flash Analog to Digital
Converters (FADCs): these are called the `high-gain' and `low-gain'
channels in what follows. The responses of the detectors are calibrated
in units of the signal produced by a muon traversing the water
vertically at the center of the station: this unit is termed the
``Vertical Equivalent Muon'' or VEM \cite{r9}. Air showers are
identified using a 3-fold coincidence, satisfied when a triangle of
neighbouring stations is triggered \cite{r10}. Using the six
FADCs, 768 samples (over 19.2 $\mu$s) are recorded at each triggered
station. For the present analysis events are used that are confined
within the array so that an accurate reconstruction is ensured. This
requires that all six stations of the hexagon surrounding the detector
with the highest signal are operational and is known as the 6T5
condition. The arrival directions are found from the relative arrival
times of the shower front at the triggered stations. The angular
resolution is 0.8\textsuperscript{o} for energies above 3 EeV for the 1500 m array and 1\textsuperscript{o}
for the 750 m-configuration \cite{r8}.

The estimator of the primary energy is the signal reconstructed at 1000
m (1500 m array) or 450 m (750 m array) from the shower core, denoted by
\textit{S}(1000) and \textit{S}(450) respectively. These estimators are
determined, together with the core position, through a fit of the
recorded signals (converted to units of VEM after integration of the
FADC traces) to a lateral distribution function that describes the
average rate of fall-off of the signals as a function of the distance
from the shower core. For \textit{S}(1000) \textgreater{} 20 VEM
(corresponding to an energy of $\sim$3 EeV) showers are
recorded with full efficiency over the whole area of the array. For the
750 m array, the corresponding value of \textit{S}(450) for full
efficiency is $\sim$60 VEM ($\sim$0.5 EeV). The
accuracy of the core location in lateral distance is $\sim$50 m (35 m) for the
two configurations. The uncertainty of \textit{S}(1000), which is
insensitive to the lateral distribution function \cite{r11}, is 12\% (3\%) at 3
EeV (10 EeV) and for \textit{S}(450) the corresponding figure is 30\% at
0.5 EeV.

The conversions from \textit{S}(450) and \textit{S}(1000) to energy are
derived using subsets of showers that trigger the fluorescence detector
and the surface array independently (`hybrid events') using
well-established methods \cite{r12}. The statistical uncertainty
in the energy determination is about 16\% (12\%) for the two reference
energies of the 1500 m array and 15\% at 0.5 EeV for the 750 m array.
The absolute energy scale, determined using the fluorescence detector,
has a systematic uncertainty of 14\% \cite{r13}.

Twenty-four telescopes (each with a field of view of
30\textsuperscript{o}$\times$30\textsuperscript{o}) form the Fluorescence
Detector (FD). They are distributed in sets of six at four observation
sites. The FD overlooks the SD array and collects the fluorescence light
produced as the shower develops in the atmosphere. Its duty cycle
amounts to $\sim$15\% since it operates exclusively on clear
moonless nights. The 750 m array is overlooked from one observation site
by three high-elevation telescopes (HEAT) with a field of view covering
elevations from 30\textsuperscript{o} to 60\textsuperscript{o}, thus
allowing the study of lower energy showers. Those low energy showers are detected 
closer to the detector; therefore we need a higher elevation field of view to contain 
also this kind of events.

\section{The Risetime and its Measurement}
\label{s3}

\subsection{Overview of the risetime concept}
\label{s3.1}

In the study described below, we use the risetime of the signals from the water-Cherenkov detectors to extract information about the development of showers. A single parameter, namely the time for the signal to increase from 10\% to 50\% of the final magnitude of the integrated signal, \textit{t}\textsubscript{1/2}, is used to characterize the signal at each station. This parameter was used for the first demonstration of the existence of ‘between-shower’ fluctuations in an early attempt to get information about the mass of the cosmic rays at $\sim$1 EeV  \cite{r14}. That work was carried out using data from the Haverah Park array in England where the water-Cherenkov detectors were of 1.2 m deep (identical to those of the Auger Observatory) and of area 34 m\textsuperscript{2}. 

\begin{figure}
\centering
\includegraphics[width=\columnwidth]{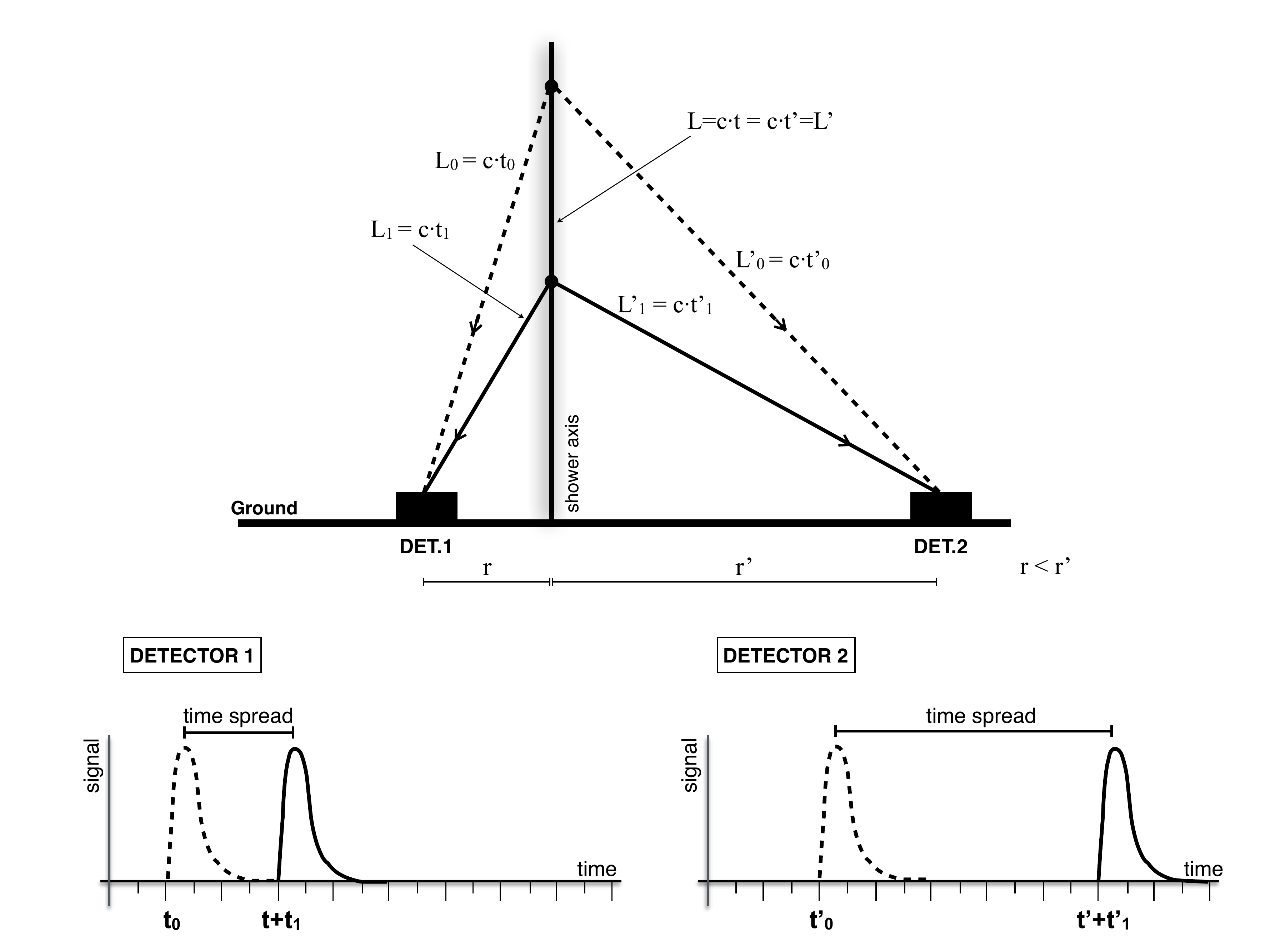}
\caption{Qualitative sketch of how geometrical effects affect the temporal spread of the muons at a detector.}
\label{fig01}
\end{figure}

\begin{figure*}
\centering

\begin{tabular}{cc}
\subfigure{\includegraphics[width=80mm]{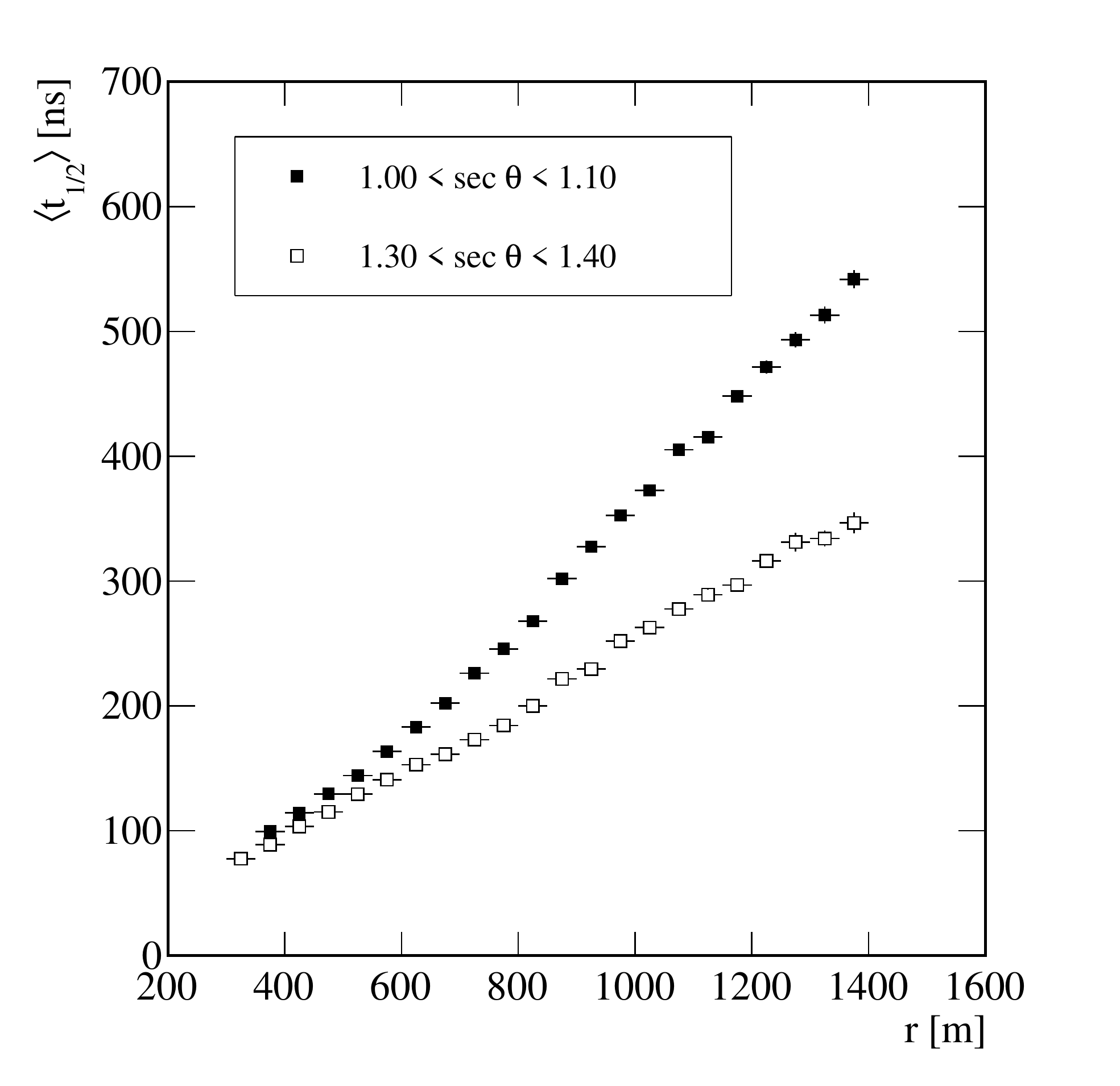}}
&
\subfigure{\includegraphics[width=80mm]{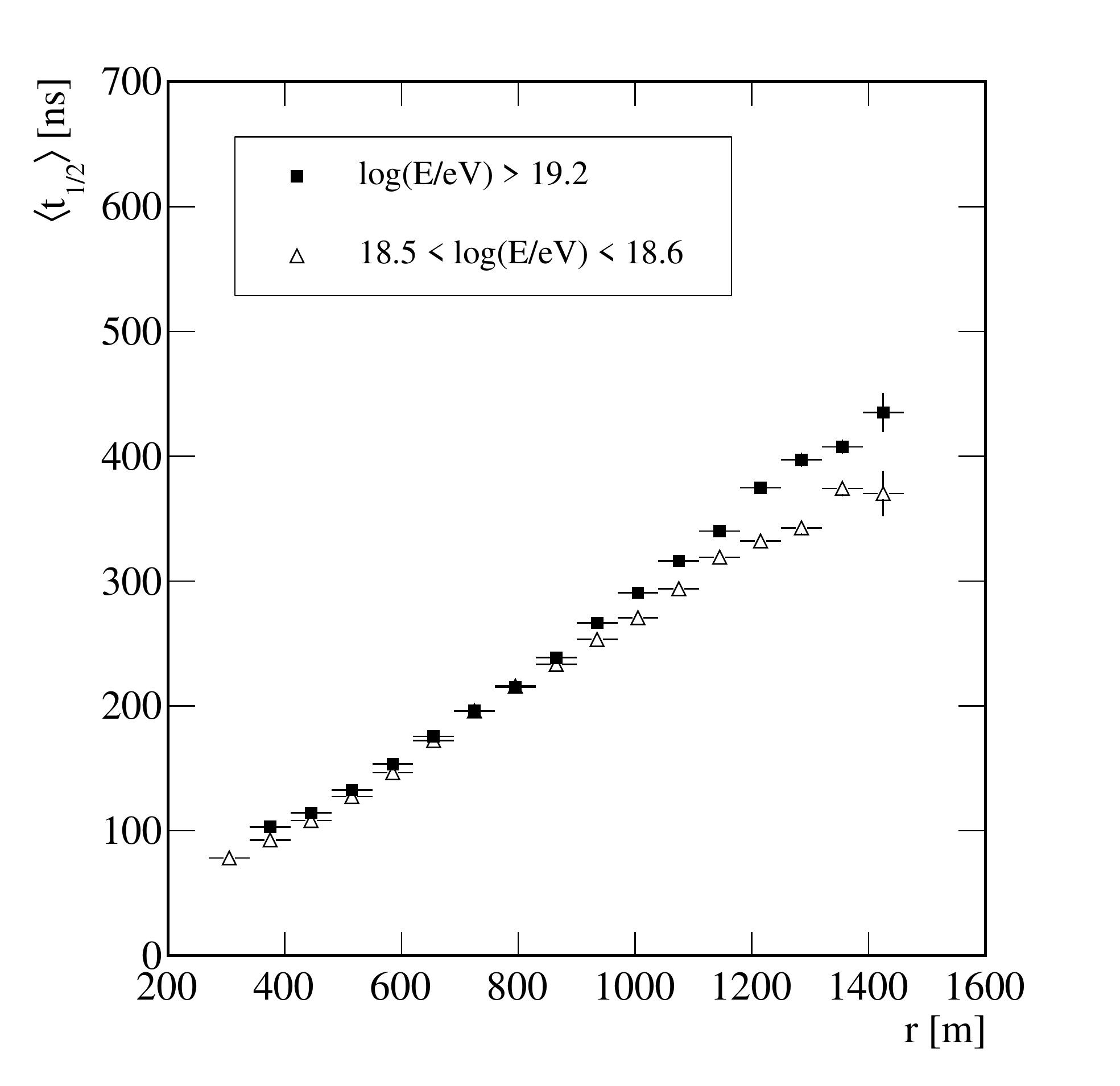}}
\\
\subfigure{\includegraphics[width=80mm]{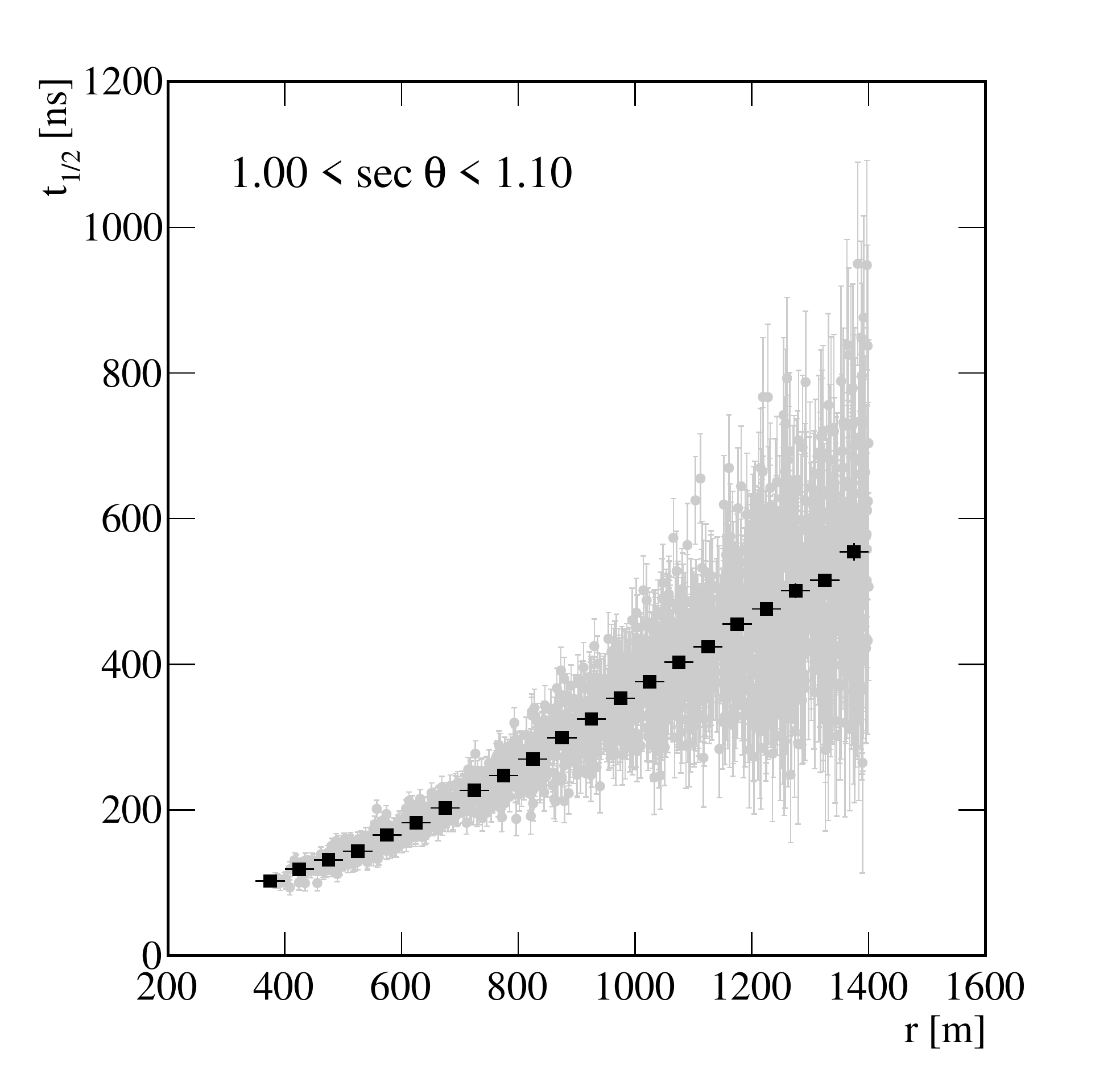}}
&
\subfigure{\includegraphics[width=80mm]{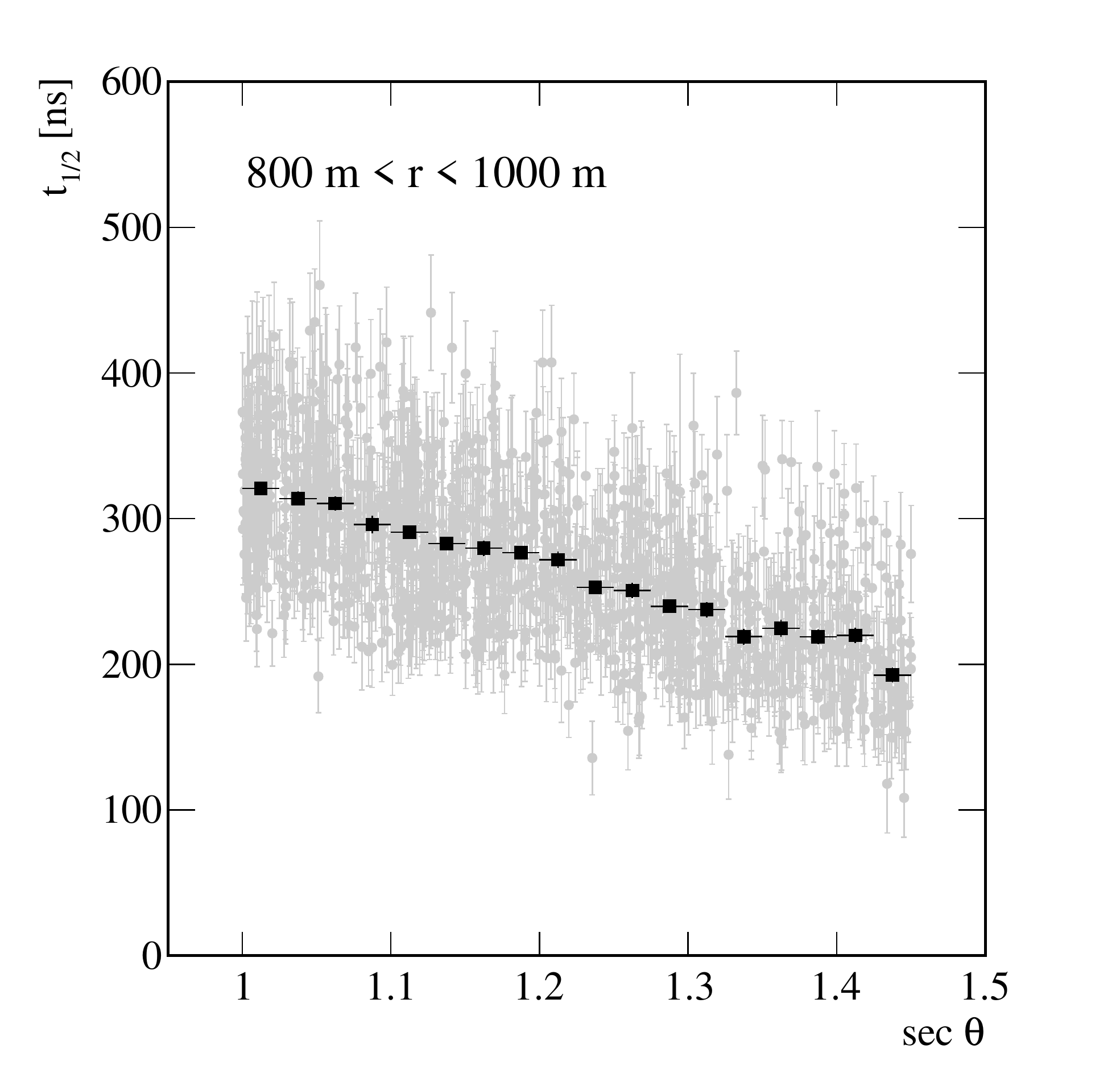}}
\\
\end{tabular}
\caption{ (Top left) The risetime as a
function of distance to the shower core for two different intervals of
sec $\theta$ in the energy range 19.0 \textless{} log (\textit{E}/eV)
\textless{} 19.2. (Top right) The risetime as a function of distance for
two different energy bands in the angular range 1.20 \textless{} sec
$\theta$ \textless{} 1.30. (Bottom left) Illustration of the spread in
the risetimes as a function of distance for events in the energy range
19.1 \textless{} log (\textit{E}/eV) \textless{} 19.2. (Bottom right)
Illustration of the variation of risetime with zenith angle for events
in the energy range 19.1 \textless{} log (\textit{E}/eV) \textless{} 19.2. 
All plots are based on experimental data}
\label{fig:MPDvsr}
\end{figure*}

The choice of this parameter is based on the experimental work by Linsley and Scarsi \cite{r15}. They showed that at distances of more than about 100 m from the shower core, the early part of the shower signal is dominated by muons. Direct measurements of muons using magnetic spectrographs established that the mean momentum of muons beyond 100 m was more than 1 GeV: this leads to the conclusion that the geometrical effects dominate the temporal spread of the muons at a detector.  This is illustrated in Fig. \ref{fig01} for a shower arriving in the vertical direction where it can be seen that the muons arriving from lower down in the shower arrive later at a detector than those that arise from higher up.  Furthermore it is evident that at larger distances from the shower axis, the muons will be more dispersed in time than at smaller distances, leading to the dependence of the risetime on distance found experimentally (Fig. \ref{fig:MPDvsr}).
Because the muons are relatively energetic, the effects of velocity difference, of deflections in the geomagnetic field and of Coulomb scattering is small although these factors were taken into account even in the earliest Monte Carlo studies of the phenomenon \cite{r16}.  By contrast the electrons and photons of an air shower have mean energies of about 10 MeV so that the arrival of the electromagnetic component of the shower is delayed with respect to the muons because of the multiple scattering of the electrons.  The delay of the electromagnetic component with respect to the muons also increases with distance.

The risetime is found experimentally to be a function of distance, zenith angle and energy (Fig. \ref{fig:MPDvsr}). At 1000 m from the shower axis, for a vertical event of 10 EeV, \textit{t}\textsubscript{1/2} $\sim$380 ns. This value increases slowly with energy and decreases with zenith angle. At large angles and/or small distances \textit{t}\textsubscript{1/2} can be comparable to the 25 ns resolution of the FADCs and this fact restricts the data that are used below. The fastest risetime, measured in very inclined showers or with single muons, is 40 ns and is an indication of the limitations set by the measurement technique and hence guides our selection of distance and angular ranges.

Because of the size of the Auger Observatory and the large separation of the detectors, it is necessary to take account of the fact that a detector that is struck early in the passage of the shower across the array will have a slower risetime than one that is struck later, even if the two detectors are at the same axial distance from the shower core. This asymmetry arises from a complex combination of attenuation of the electromagnetic component as the shower develops and because of the different part of the angular distribution of muons (more strictly of the parent pions) that is sampled at different positions across the array. The attenuation of the electromagnetic component of a shower across an array was first discussed by Greisen \cite{r17}. A detailed description of the asymmetry that is observed, and of its power for testing hadronic interaction models, has been given recently \cite{r6}. For the present study, the asymmetry is taken into account by referencing each risetime to that which would be recorded at a hypothetical detector situated at 90\textsuperscript{o} with respect to the direction of the shower axis projected onto the ground, and at the same axial distance from the shower core, as the station at which a measurement is made. The amplitude of the asymmetry is a function of zenith angle, axial distance and energy: at 40\textsuperscript{o}, 750 m and 10 EeV it is $\sim$15\%.

The magnitudes of the risetimes that are measured in a particular shower depend upon the development of the shower.  As the energy increases, the mean position of the point of maximum development of the shower moves deeper into the atmosphere and thus the risetimes will, on average, be slower than for a lower energy event.  Because muons dominate the shower to an increasing extent at large zenith angles, because the electromagnetic component suffers increased attenuation, the risetimes are expected to be faster at a detector that is at the same distance from the axis of the shower but in the vertical direction.  The magnitude of the energy, distance and zenith angle effects that can be inferred qualitatively from Fig. \ref{fig01} are evident in the data shown in Fig. \ref{fig:MPDvsr}.

From these considerations, it follows that studying the risetimes of showers provides a method of measuring the shower development and thus of deducing the mass composition.  Details of the study are presented below where the risetime properties are also compared with predictions from Monte Carlo calculations using different hadronic models.

\subsection{Determination of the accuracy of measurements of
\textit{\MakeLowercase{t}}\textsubscript{1/2}}

The uncertainty in a measurement of \textit{t}\textsubscript{1/2} is found
empirically from the data and will be described in some detail as it
plays an essential role in the determination of the new parameter,
introduced in section \ref{s4}, used to characterize shower development. The
uncertainty can be obtained by using sets of detectors placed 11 m apart
(known as `twins') and also by using detectors that lie at similar
distances from the shower core (`pairs'). Measurements made using twins
and pairs cover different distance ranges. With twins we can
parameterize the uncertainty with a sufficient number of events only
between 300 m and 1200 m from the shower core. With pairs we can cover
distances from 600 m to 1800 m. It is then natural to combine both sets
of measurements to avoid as much as possible relying on extrapolations
when estimating the uncertainty in the measurement of
\textit{t}\textsubscript{1/2}.

The twin detectors give two independent measurements of the risetime at
what is effectively a single point in the shower plane. Differences in
the values of \textit{t}\textsubscript{1/2} at the twins arise from the
limitations set by the sampling of the shower front by a detector of
finite size (10 m\textsuperscript{2}) and from the measurement
uncertainties intrinsic to the FADC system. For the more-numerous pairs
there are the additional complications that arise from the asymmetry
effect and from the difference in distance of the pairs from the shower
core.

\subsubsection{Assessment of measurement uncertainty using twin
detectors}

In the surface-detector array there are 14 sets of twins and seven sets
of triplets (three detectors on a triangular grid each separated by 11
m): the triplets are also referred to as `twins'. We parameterize the
uncertainty by splitting the data in different bins of distance to the
core, zenith angle and detector signal. This implies that a precise
parameterization of the uncertainty demands a large amount of data. To
cope with this requirement we must combine all twin measurements that
belong to events reconstructed at either of the arrays. A total of
$\sim$83 000 twin measurements are available from the two
arrays for zenith angles below 60\textsuperscript{o} and above energies of 0.3 EeV and 1
EeV for events that trigger the two arrays. The cuts on energy and
zenith angle are very loose to enhance the number of events available
for analysis. Likewise the criteria applied at detector level and
detailed in Table \ref{t1} are mild to keep the selection efficiency as high as
possible. We discard detectors that recorded a small number of particles
or located far from the core to avoid biases in the signal measurement.
For very large signals, the risetime measurements approach the
instrumental resolution and therefore are discarded. The cut on
\textbar{}\textit{S}\textsubscript{i} --
\textit{S}\textsubscript{mean}\textbar{} in Table \ref{t1} is made to deal with cases
where one signal is typically around 5 VEM and the other, possibly
because of an upward fluctuation, is relatively large. Such twins are
rejected.

The average uncertainty in a risetime measurement,
$\sigma$\textsubscript{1/2,} is given by
\begin{equation}
\sigma _{1/2} =\frac{\sqrt{\pi } }{2} \left\langle \left| t_{1/2}^{1} -t_{1/2}^{2} \right| \right\rangle
\end{equation}
where the superscripts define each member of the twin. As twin detectors
are only 11 m apart no correction is necessary for the azimuthal
asymmetry.

\begin{table*}
\caption{Selection of twin detectors 
used to assess the risetime uncertainty.}
\centering 
\begin{tabular}{lcc||cc}
\hline\hline
& \multicolumn{2}{c}{750 m array} & \multicolumn{2}{c}{1500 m
                                    array} \\ \hline\hline 
Cuts & Number of twins & Efficiency & Number of twins &
Efficiency \\ \hline
Pre-twin selection & 41 100 & 1.00 & 41 934 & 1.00\\
5 \textless{} \textit{S}/VEM \textless{} 800 & 34 461 & 0.84 & 35 704 &
0.85\\
\textit{r} \textless{} 2000 m & 34 459 & 0.83 & 35 620 &
0.84\\
\textbar{}\textit{S}\textsubscript{i}-\textit{S}\textsubscript{mean}\textbar{}\textless{}
0.25 \textit{S}\textsubscript{mean} & \textbf{28 466} & 0.69 & \textbf{29
832} & 0.71 \\  \hline\hline
\end{tabular}
\label{t1}
\end{table*}

The data have been divided into seven sec $\theta$ intervals (of width
0.1) and six distance ranges (see Fig. \ref{fig3} left). The mean values of
$\sigma$\textsubscript{1/2} as a function of signal size, \textit{S}, are
fitted with the function

\begin{equation}
\sigma _{1/2} \, =\sqrt{\left( \frac{J(r,\theta )}{\sqrt{S} } \right) ^{2} +\left( \sqrt{2} \frac{25\, \text{ns} }{\sqrt{12} } \right) ^{2} }
\end{equation}

The first term in this function represents the differences seen between
the two detectors while the second term arises from the digitisation of
the signal in time intervals of 25 ns.

\textit{J} is found from a linear function, \textit{J}(r,$\theta$) =
\textit{p}\textsubscript{o}($\theta$) + \textit{p}\textsubscript{1}($\theta$)
r, and the fitted values of \textit{p}\textsubscript{o} and
\textit{p}\textsubscript{1} as functions of sec $\theta$ are
\begin{equation}
\begin{array}{l}
p_{0} (\theta )=(-340\pm 30)+(186\pm 20)\, \sec \theta  \\
p_{1} (\theta )=(0.94\pm 0.03)+(-0.44\pm 0.01)\, \sec \theta \end{array}
\end{equation}
where \textit{p}\textsubscript{0} units are (ns VEM\textsuperscript{1/2})
and \textit{p}\textsubscript{1} units are (ns VEM\textsuperscript{1/2}
m\textsuperscript{-1}).

\subsubsection{Assessment of measurement uncertainty using pairs of
detectors}

For the purposes of this study, a pair of detectors is defined as any
two detectors in the same shower where the difference in distance from
the shower core, (\textbar{}\textit{r}\textsubscript{2} --
\textit{r}\textsubscript{1}\textbar{}) is less than 100 m, irrespective of
azimuth angle. After applying the cuts previously discussed, there are
$\sim$50\% more pair measurements than there are measurements
from twins. This sample is large enough to allow us to limit this study
to pairs of detectors from the 1500 m array only. However, corrections
have to be made for the asymmetry and, because of the array spacing,
there are no data below 600 m. Additionally a correction must be made
because the risetimes are at different axial distances: for a 100 m
separation this difference is $\sim$30 ns, assuming a linear
dependence of risetime with distance (see Fig. \ref{fig:MPDvsr}). Before applying this
correction the mean time difference for pairs was (14.750 $\pm$ 0.002) ns:
after correction the average difference was (0.140 $\pm$ 0.002) ns.

\begin{figure}
\centering
\includegraphics[width=\columnwidth]{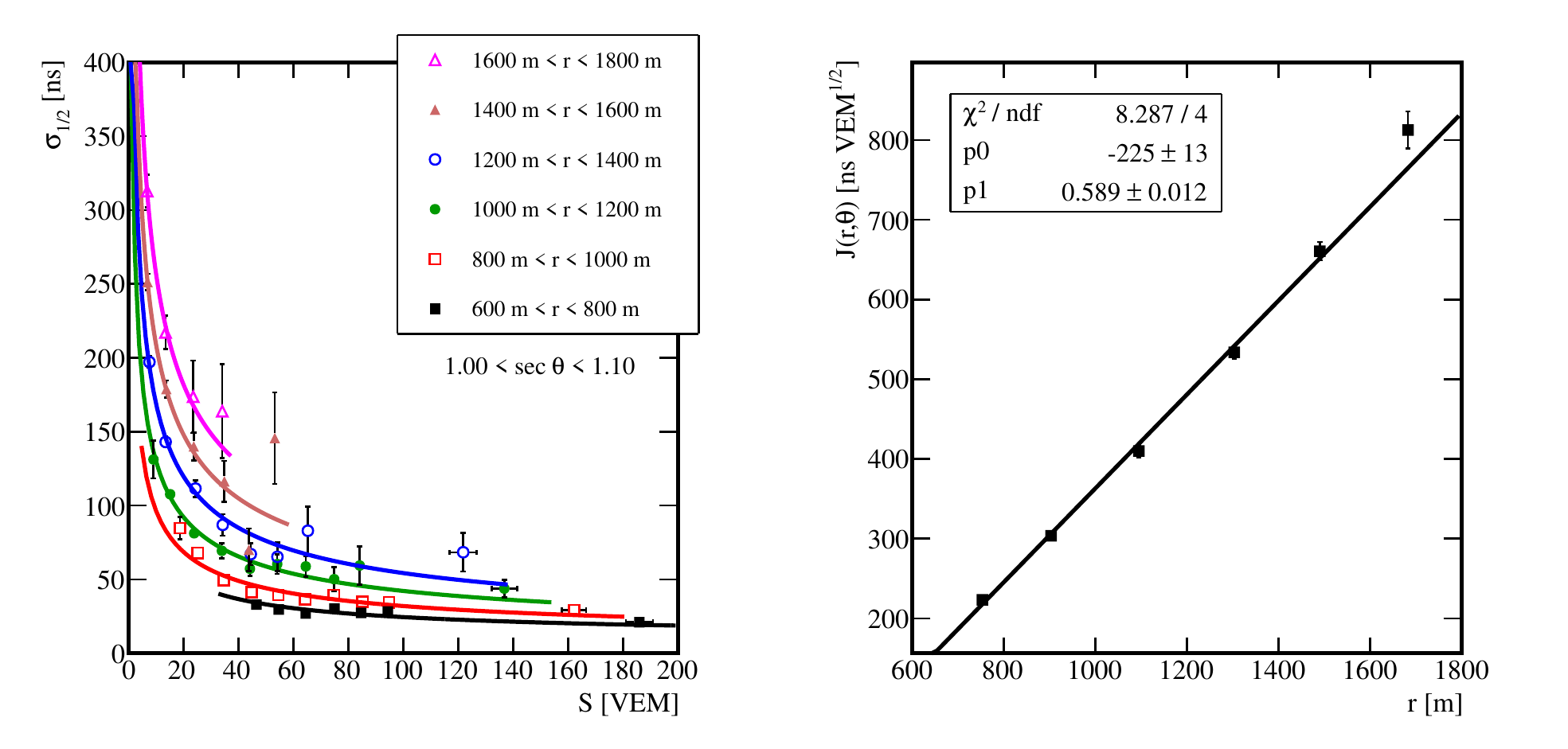}
\caption{ (Left) Uncertainty obtained
with pair detectors as a function of the station signal for vertical
events (1.00\textless{} sec $\theta$ \textless{}1.10). Each line is the
result of the fits performed for different distance ranges. Each point
represents the average measurements of at least 10 pair detectors.
(Right) The parameter \textit{J}(\textit{r}, $\theta$) as a function of the
distance to the core for the same zenith angle range.}
\label{fig3}
\end{figure}

From a similar analysis to that described for the twin detectors, the
fits for \textit{p}\textsubscript{0} and \textit{p}\textsubscript{1} have
the parameterisations

\begin{equation}
\begin{array}{l}
p_{0} (\theta )=(-447\pm 30)+(224\pm 20)\, \sec \theta  \\
p_{1} (\theta )=(1.12\pm 0.03)+(-0.51\pm 0.02)\, \sec \theta .\end{array}
\end{equation}

The variation of \textit{J} with distance is also shown in Fig. \ref{fig3}.

The differences in the values of \textit{p}\textsubscript{0} and
\textit{p}\textsubscript{1} from the two analyses arise because they cover
different distance ranges and different energy ranges. To optimize the
determination of $\sigma$\textsubscript{1/2} for the risetimes measured
at each station, we adopt the following parameterisations for
\textit{p}\textsubscript{0} and \textit{p}\textsubscript{1} for different
core ranges

\begin{equation}
\begin{array}{l} 
p_{0} (\theta ) = 
\begin{cases}
(-340\pm 30)+ (186\pm 20)\, \sec \theta,\, \, r\leq \, 650\, \text{m}
\\
(-447\pm 30)+ (224\pm 20)\, \sec \theta,\, \, r > \, 650\, \text{m} \\
\end{cases} \\
p_{1} (\theta ) = 
\begin{cases}
(0.94\pm 0.03)+ (-0.44\pm 0.01)\, \sec \theta,\, \, r\leq \, 650\, \text{m}
\\
(1.12\pm 0.03)+ (-0.51\pm 0.02)\, \sec \theta,\, \, r > \, 650\, \text{m} \\
\end{cases}
\end{array}
\end{equation}

We have set the break point at 650 m because at this distance the
uncertainties given by the two parameterizations agree within their
statistical uncertainties (2-3 ns).

\section{The new parameter $\Delta_s$ and its determination for individual air-showers}
\label{s4}

\subsection{Introduction to the Delta method}

\begin{figure}
\centering
\includegraphics[width=\columnwidth]{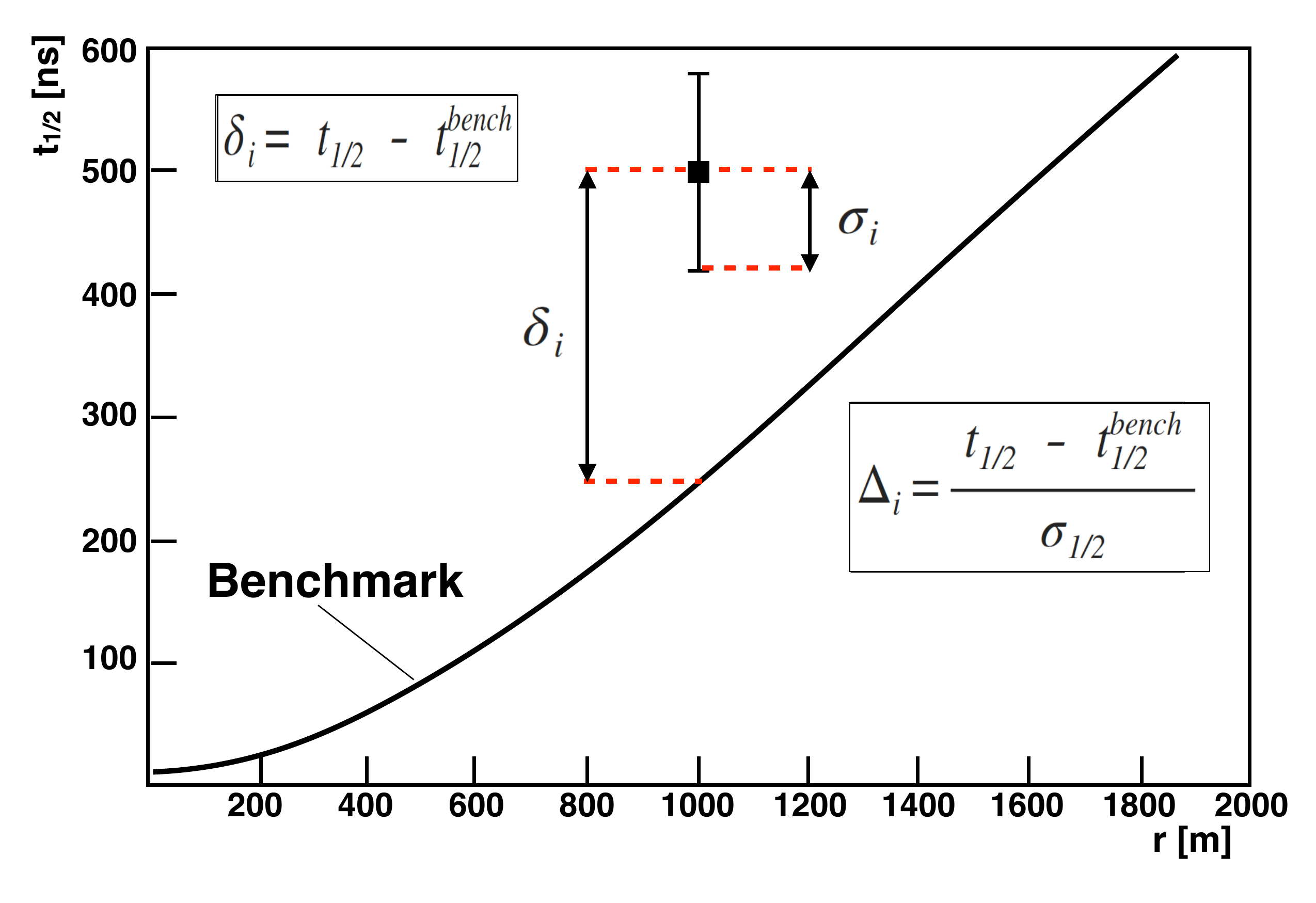}
\caption{ Schematic diagram to
illustrate the Delta method.}
\label{fig:cartoon}
\end{figure}

When a large number of risetimes is recorded in an event, it is possible
to characterize that event by a single time just as the size of an event
is designated by \textit{S}(1000), the signal size at 1000 m from the
shower axis. This approach is only practical at high energies, as
several measurements are needed to estimate the risetime at 1000 m by
extrapolation \cite{r18}. Here, to obtain a large sample of data over a wide range
of energies, an alternative method of using risetime measurements is
introduced. We have determined for the two arrays independent
relationships that describe the risetimes as a function of distance in a
narrow range of energy (see section \ref{s4.3}). We call these functions
`benchmarks', and risetimes at particular stations, after correction for
the asymmetry effect, are compared with the relevant times from the
benchmark, $t_\text{1/2}^\text{bench}$, in units of the accuracy with which they are
determined. The approach is illustrated in Fig. \ref{fig:cartoon}: the benchmarks are, of
course, zenith-angle dependent (see Fig. \ref{fig:MPDvsr}). We use the term `Delta method' to
refer to this approach in what follows.

Thus for each measurement of \textit{t}\textsubscript{1/2} at a single
detector, \textit{i}, an estimate of $\Delta_i =
\left(t_\text{1/2} - t_\text{1/2}^\text{bench}\right)/\sigma_\text{1/2}$ is
made. Each shower is then characterised by $\Delta_s$,
the average of these estimates for the \textit{N} selected stations.

\begin{equation}
\Delta _{s} =\frac{1}{N} \sum\limits_{stations}^{}\Delta _{i} 
\end{equation}

\subsection{Data selection}

The data from the water-Cherenkov arrays were collected between 1 Jan
2004 (2008 for the 750 m spacing) and 31 Dec 2014. The first selection,
of 6T5 events, has already been discussed. Other selections for the two
arrays are shown in Table \ref{t2}.

\begin{table*}
\caption{ Quality cuts applied to the
events of the 750 m and the 1500 m arrays. $\epsilon$ stands for the
overall efficiency. The explanation for the different cuts can be found
in the text.}
\centering
\begin{tabular}{lrr||lrr}
\hline\hline
 \multicolumn{3}{c}{750 m array} & \multicolumn{3}{c}{1500 m array} \\
  \hline \hline 
\multicolumn{1}{l}{Quality cuts} & \multicolumn{1}{c}{Events} &
                                                                \multicolumn{1}{c||}{$\epsilon$
                                                                (\%)
                                                                } & \multicolumn{1}{l}{Quality cuts} & \multicolumn{1}{c}{Events} & \multicolumn{1}{c}{$\epsilon$ (\%}) \\
17.5 \textless{} log (\textit{E}/eV) \textless{} 18.5 & 159 795 & 100.0 &
log (\textit{E}/eV) \textgreater{} 18.5 & 217 469 & 100.0\tabularnewline
sec $\theta$ \textless{} 1.30 & 72 907 & 45.6 & sec $\theta$ \textless{}
1.45 & 97 981 & 45.0\\
6T5 trigger & 29 848 & 18.7 & 6T5 trigger & 67 764 & 31.0\tabularnewline
Reject bad periods & 28 773 & 18.0 & Reject bad periods & 63 856 &
29.0\\
$\geq$ 3 selected stations & \textbf{27 553} & \textbf{17.2} & $\geq$ 3 selected
stations & \textbf{54 022} & \textbf{24.8}\\
\hline\hline
\end{tabular}
\label{t2}
\end{table*}

The lower energy cuts are made to select events that trigger the arrays
with 100\% efficiency. The upper energy cut in the 750 m array is made
to set aside events in overlapping energy regions that will be used
later to cross-check the robustness of the method. As previously
discussed, at large angles \textit{t}\textsubscript{1/2} can be comparable
to the 25 ns resolution of the FADCs and this fact restricts the usable
angular range. The cut in zenith angle is lower for the 750 m array than
for the 1500 m array because the stations tend to be closer to the core
and the limitations set by the sampling speed of the FADCs become more
important at larger angles and small distances. We rejected data taking
periods where the performance of the array of surface detectors was not
optimal. At least three selected stations are required for an event to
be included in the data samples.

The stations used within each event must fulfil the following criteria.
The stations cannot be saturated in the low-gain channel since risetimes
cannot be obtained from such signals. The signals recorded by the
stations must be bigger than 3 VEM and 5 VEM for the 750 m and the 1500
m arrays respectively. Those cuts guarantee that no bias towards
primaries of a particular type is introduced: the difference in selection efficiency 
for protons and iron is less than 5\% for all energy bins. The selected stations must
lie within a given distance range from the position of the reconstructed
core of the shower. The lower range of distance, 300 m, is selected to
avoid the problems set by the inability of the recording system to
record fast pulses (see section \ref{s3.1}), while the upper ranges (800 m
(1400 m) for the 750 m (1500 m) array) are chosen to span what is
consistent with unbiased selection. For the highest energies this has
been extended to 2000 m as the signal sizes in such events are
sufficiently large to give accurate measurements. For the 750 and 1500 m
arrays the overall selection efficiencies at station level are 52\% and
56\% respectively. This translates into 113,661 and 210,709 detectors
for the 750 m and the 1500 m arrays respectively.

Using simulations, we have searched for biases that might be introduced
into inferences about mass composition as a result of these cuts. The
difference between the overall selection efficiencies for protons and
Fe-nuclei are smaller than 2\%. The upper limit on the energy cut in the
750 m data eliminates only 2\% of the events. This cut, and the lower
energy limit for the 1500 m array, are relaxed later to study the
overlap region in detail.

For the 750 and 1500 m arrays, the mean numbers of selected stations per
event satisfying the selection criteria defined in Table \ref{t2} are 4.0 and 3.6
respectively. In the analysis discussed below, selected events are
required to have 3 or more values of $\Delta_i$, but, for
an arrival direction study, in which it is desirable to separate light
from heavy primaries, one could envisage using two stations, or even
one, to infer the state of development of the shower, albeit with more
limited accuracy.

\subsection{Determination of the benchmarks for the 750 and 1500 m
arrays}
\label{s4.3}

The determination of the benchmarks, which define the average behaviour
of the risetimes as a function of distance and zenith angle, is
fundamental to the success of the technique. Essentially the same
procedures have been adopted for both arrays. For each detector two time
traces are recorded on high-gain and low-gain channels. The risetime of
a detector is computed according to the following procedure: in the case
where no saturation occurs, the risetime is obtained from the trace
corresponding to the high-gain channel. If this channel is saturated, we
use the trace from the low-gain channel to compute the risetime. If the
low-gain signal is saturated as well, which can occur for stations close
to the core in high-energy events, that station is not selected for this
analysis. Further details of the recording procedures are given in \cite{r8}.

\begin{figure}
\centering
\includegraphics[width=\columnwidth]{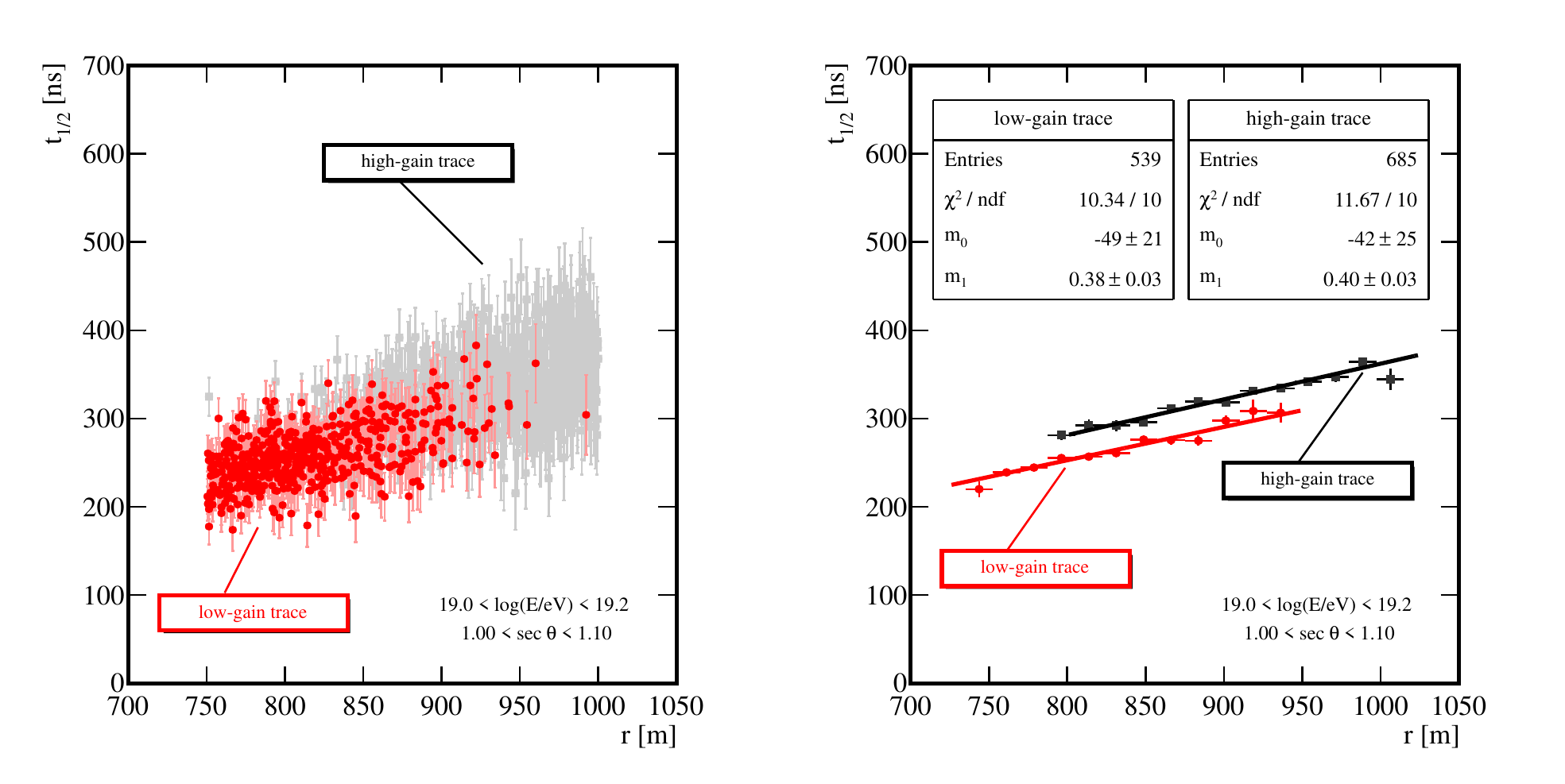}
\caption{(Left) Risetimes as a
function of core distance for events with and without saturation in the
high-gain channel. We have selected events with energies 19.0
\textless{} log (\textit{E}/eV) \textless{}19.2 and zenith angles 1.00
\textless{} sec $\theta$ \textless{}1.10. (Right) Same as left plot but
this time we show average values to make more evident the difference
between measurements when the saturation of the high-gain channel is
present.}
\label{fig4}
\end{figure}

In computing the benchmarks, account must be taken of the fact that the
risetimes measured for a station in the two channels are not identical,
as illustrated in Fig. \ref{fig4}. During the digitization process, a threshold is
imposed that removes very small signals. The net effect of this
threshold affects the low-gain traces much more, since their signals are
smaller due to the lower signal-to-noise ratio when compared to the one
associated to high-gain traces. The influence of tails in the
determination of the integrated signal is therefore reduced for low-gain
signals and as a consequence the risetime measurement is affected. This
instrumental effect makes it necessary to obtain benchmarks for the
high-gain and the low-gain traces independently.

As shown in Table \ref{t2} the energy ranges covered by the two arrays are 17.5
\textless{} log (\textit{E}/eV) \textless{} 18.5 (750 m spacing) and log
(\textit{E}/eV) \textgreater{}18.5 (1500 m spacing). The energy bins
chosen for the benchmarks of the 750 and 1500 m arrays are 17.7
\textless{} log (\textit{E}/eV) \textless{} 17.8 and 19.1 \textless{} log
(\textit{E}/eV) \textless{} 19.2 respectively. The choices for the
benchmarks are most effective in dealing with the high-gain/low-gain
problem just discussed. They guarantee that we reject a reduced number
of detectors where the low and the high-gain channels are simultaneously
saturated and therefore allow a definition of the benchmark over a broad
distance range. In addition, this implies that the distance intervals
used to fit the behaviour of the risetimes computed either with the low
or the high-gain traces are sufficiently long to avoid compromising the
quality of the fit.

A fit is first made to the data from the low-gain channels using the
relation
\begin{equation}
t_\text{1/2}^{\text{low-gain trace}}= 40 \, \text{ns} +\sqrt{A(\theta)^2+B(\theta)r^2}-A(\theta)
\end{equation}
where \textit{A} and \textit{B} are free parameters. The reason for adopting
40 ns as a limit was explained in section \ref{s3.1}. Other functions were
tested: this one gave consistently lower values of reduced
$\chi^2$ over the range of angles and energies used
for the two arrays.

Having used low-gain traces to evaluate \textit{A} and \textit{B}, the
signals from high-gain traces are now fitted with the function
\begin{equation}
t_\text{1/2}^{\text{high-gain trace}}= 40 \, \text{ns} +N(\theta)\left(\sqrt{A(\theta)^2+B(\theta)r^2}-A(\theta)\right)
\end{equation}
in which there is one free parameter, $N(\theta)$, that describes the
shift between the measurements in the two channels. Examples of the
quality of the fits of these functions to the data are shown in
Fig. \ref{fi5} and Fig. \ref{fig6} for
two angular ranges for each of the two arrays.

\begin{figure}
\centering
\includegraphics[width=\columnwidth]{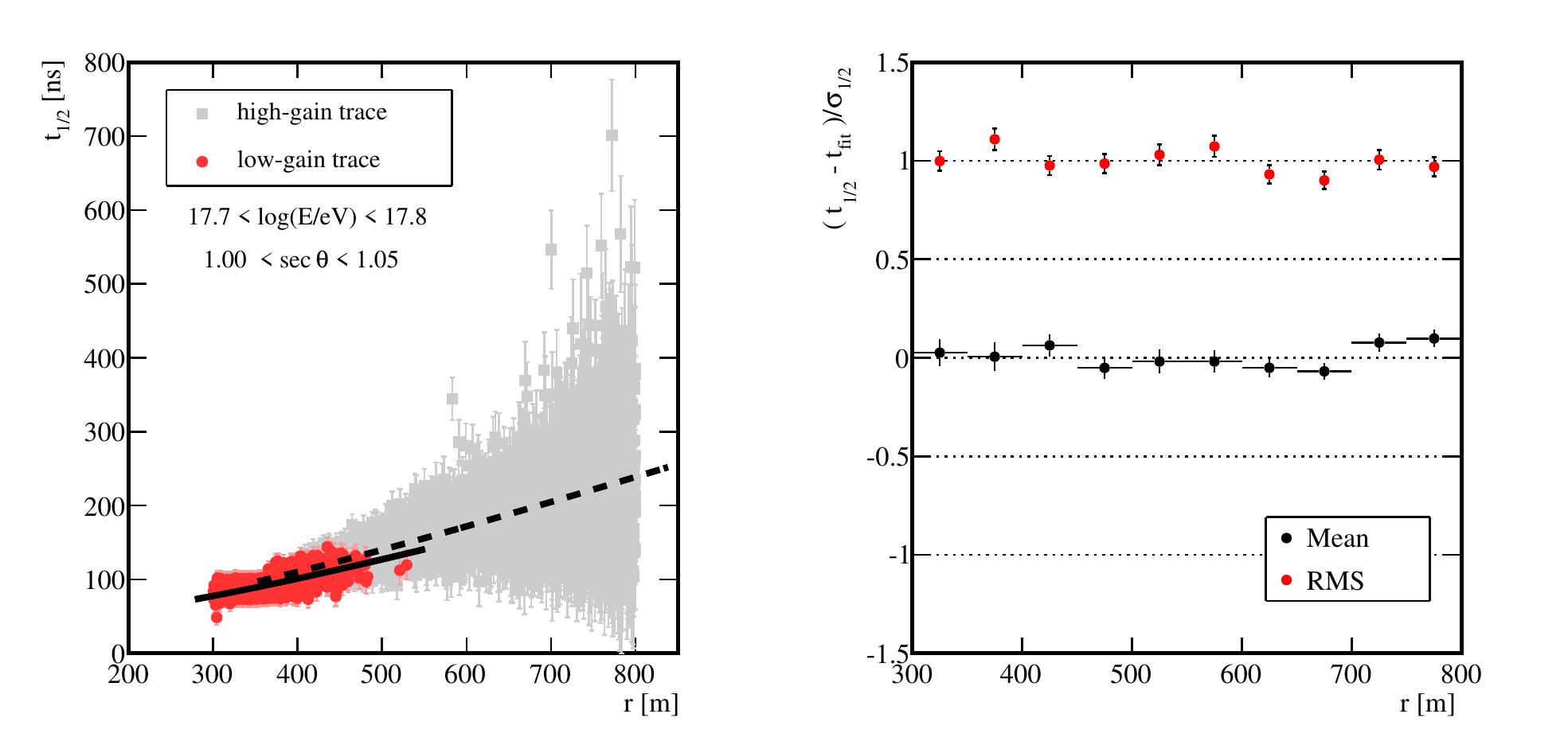}
\hfil
\includegraphics[width=\columnwidth]{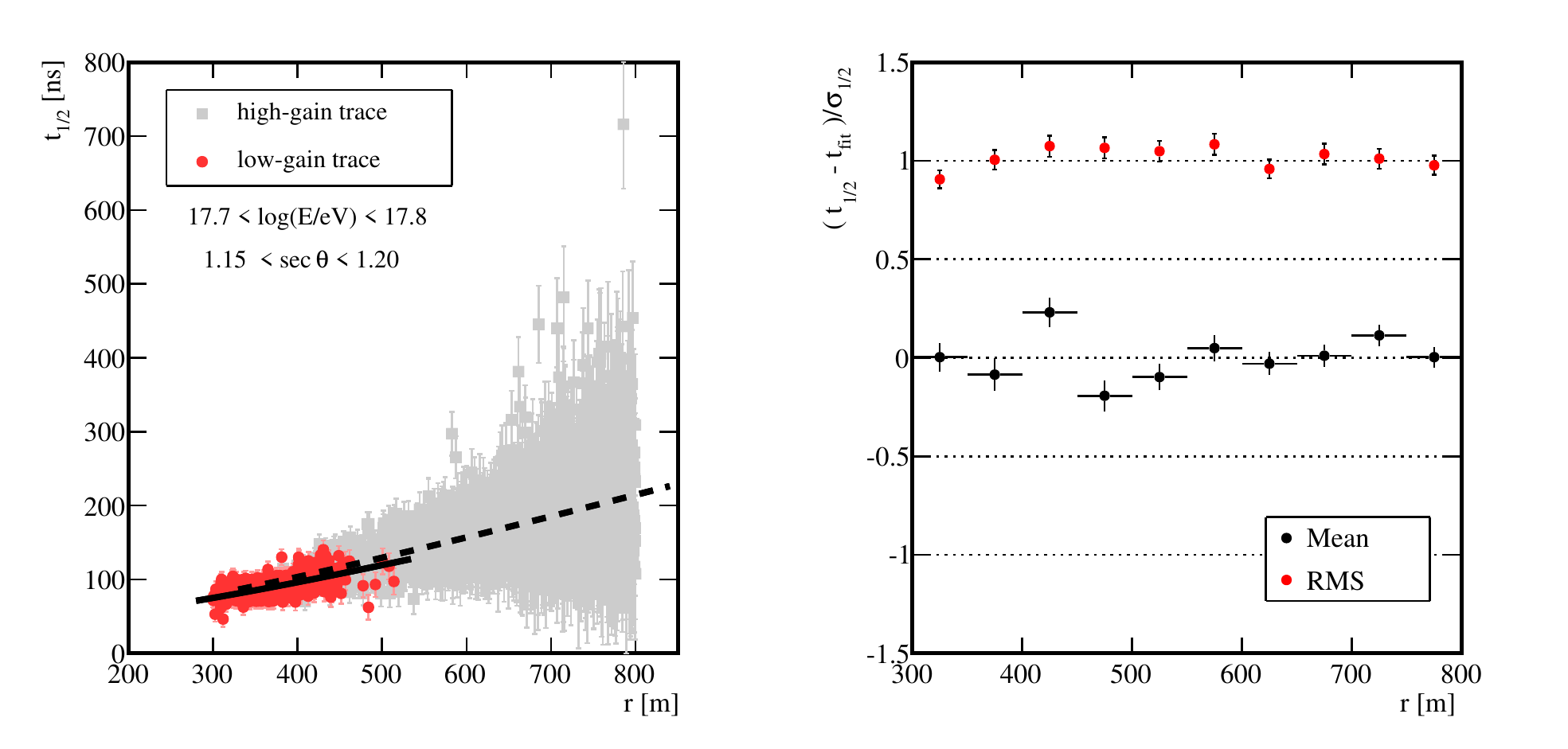}
\caption{Examples of benchmark fit
for the 750 m array. (Top panels) 1.00 \textless{} sec $\theta$
\textless{} 1.05. (Bottom panels) 1.15 \textless{} sec $\theta$
\textless{} 1.20. The solid (dashed) line corresponds to the fit done to
the risetimes computed using the low-gain (high-gain) trace.}
\label{fig5}
\end{figure}

\begin{figure}
\centering
\includegraphics[width=\columnwidth]{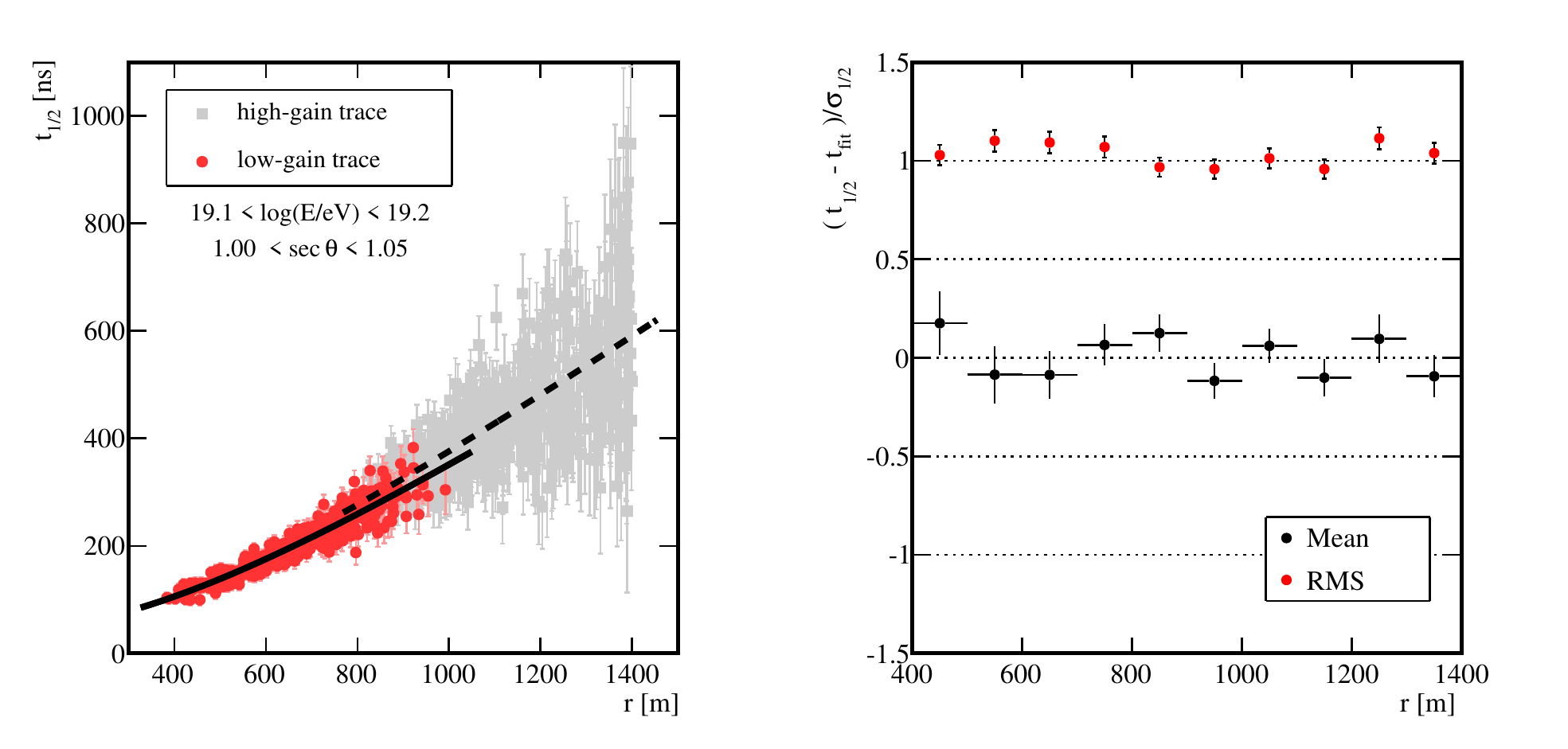}
\hfil
\includegraphics[width=\columnwidth]{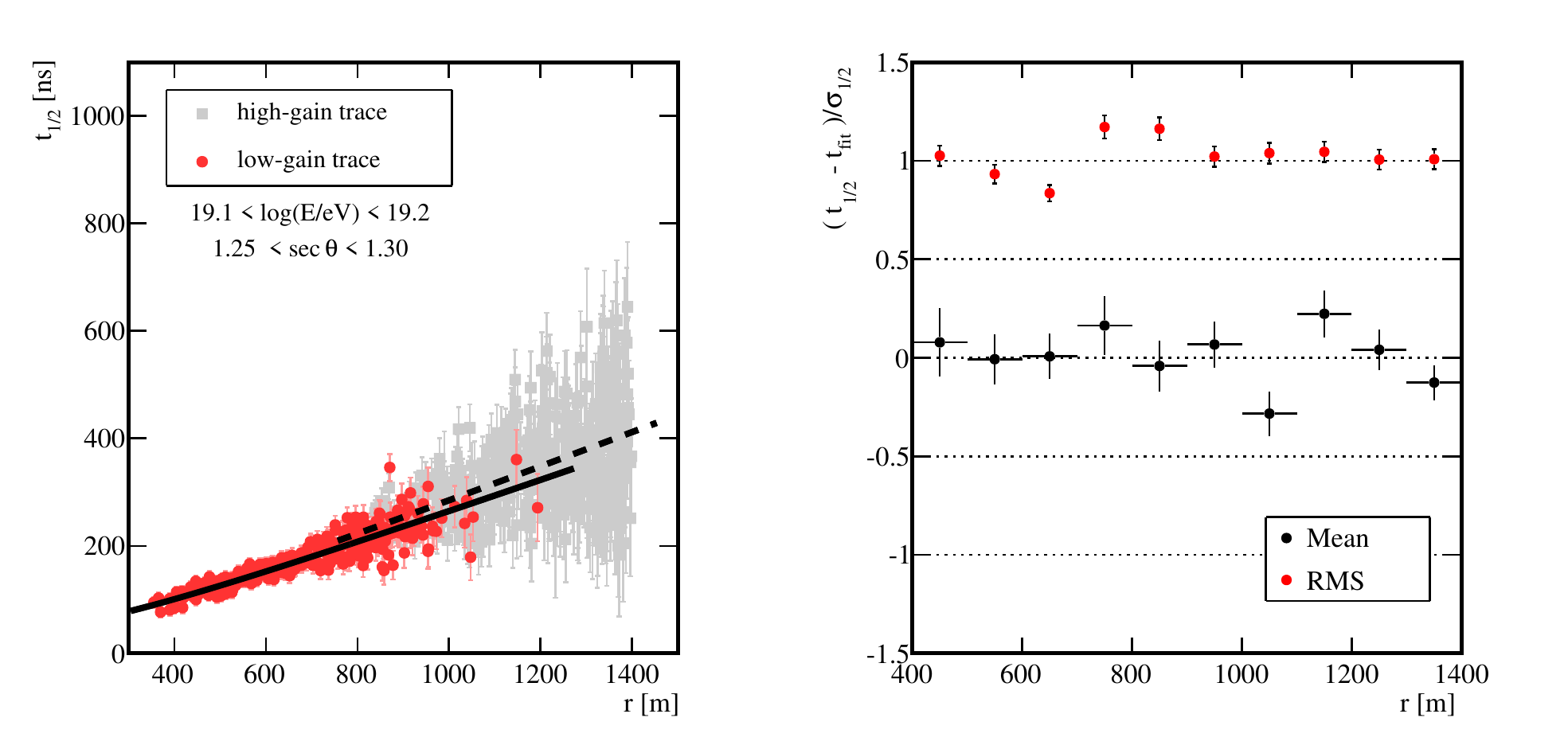}
\caption{Examples of benchmark fit
for the 1500 m array. (Top panels) 1.00 \textless{} sec $\theta$
\textless{} 1.05. (Bottom panels) 1.25 \textless{} sec $\theta$
\textless{} 1.30. The solid (dashed) line corresponds to the fit done to
the risetimes computed using the low-gain (high-gain) trace.}
\label{fig6}
\end{figure}

In the right-hand plots of each pair, the mean and RMS deviations of the
fits are seen to be consistent with 0 and 1, as expected for pull
distributions \cite{r17}. The uncertainty in the axial distance has not been
included in the fits as it is only around 2\% for the distances in
question.

Fits were made for \textit{A}, \textit{B} and $N(\theta)$ in six intervals
of sec $\theta$ ranges 1.0 -- 1.30 and 1.0 -- 1.45 for the 750 m and
1500 m arrays respectively. In all six cases the
$\chi^2$-values of the fits are between 1 and 1.2. To
obtain the final parameterization of \textit{A}, \textit{B} and \textit{N} as
a function of $\theta$ fits have been made using the following functions
\begin{equation}
\begin{array}{l}
A(\theta)= a_0+a_1 (\sec \theta)^{-4}  \\
B(\theta)= b_0+b_1 (\sec \theta)^{-4}  \\
N(\theta)= n_0+n_1 (\sec \theta)^{2}+n_2 e^{\sec \theta}
\end{array}
\end{equation}
where the seven coefficients, \textit{a}\textsubscript{0,}
\textit{a}\textsubscript{1} etc., are determined for the two arrays. This
set of functions has been empirically chosen. It guarantees that, for
the energy bins for which the benchmarks are defined, the mean value of
$\Delta_s$ shows a flat behaviour as function of sec
$\theta$. This naturally follows from the definition of  $t_\text{1/2}^\text{bench}$. Since
it embodies the dependence on sec $\theta$, the numerator in the
definition of $\Delta_i$ has to be independent of the
zenith angle.

We may thus define the benchmarks in terms of \textit{A}, \textit{B} and
\textit{N} as a function of sec $\theta$, enabling an appropriate
benchmark to be defined for the zenith angle of the event under study.
Thus $\Delta_i$ can be found for every station that
satisfies the selection criterion and the corresponding value of
$\Delta_s$ can be found for every selected event.

\section{Evolution of $\langle\Delta_s\rangle$ with energy and
comparison with model predictions}
\label{s5}

\begin{figure*}
\centering
\includegraphics[width=\columnwidth]{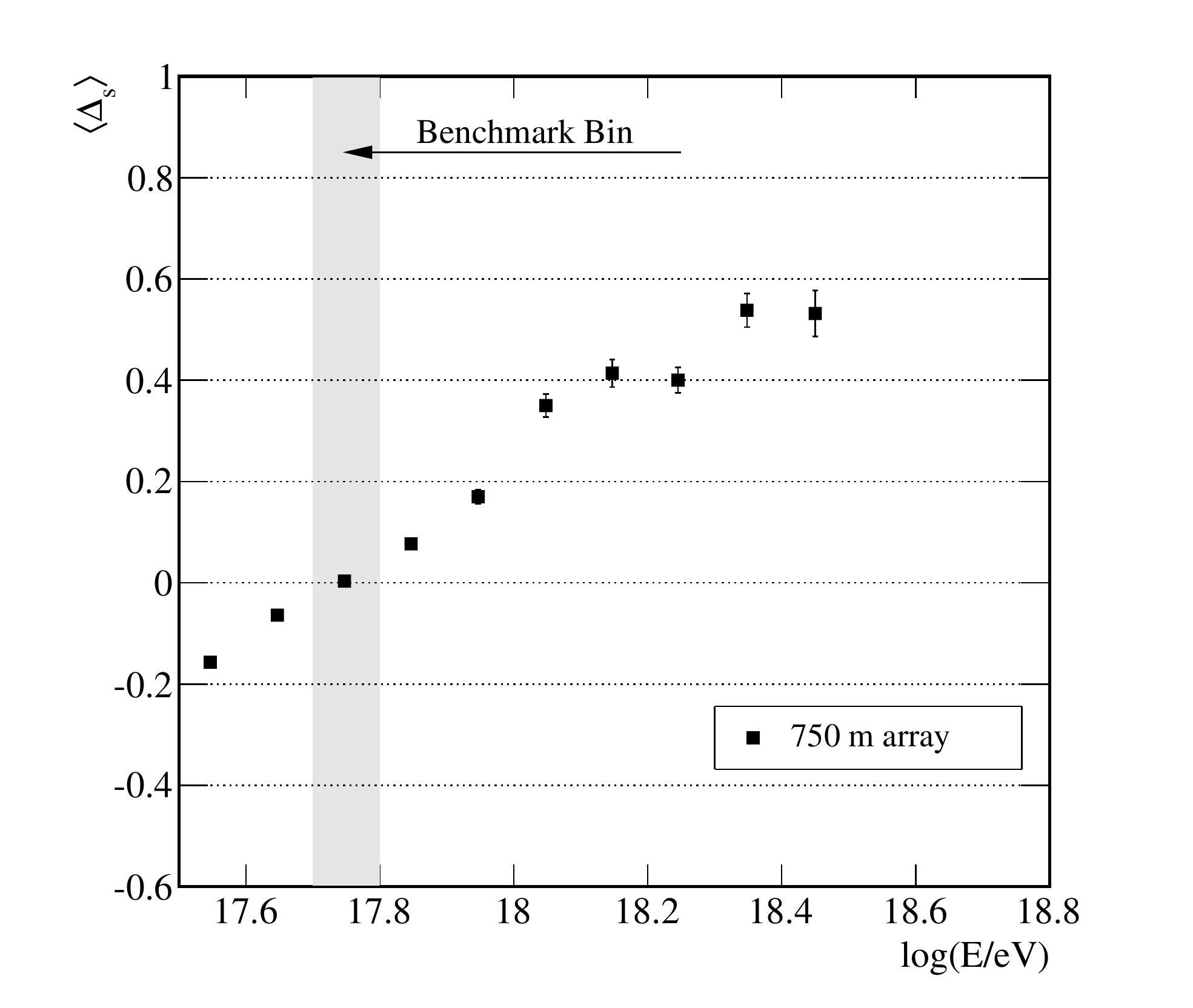}
\includegraphics[width=\columnwidth]{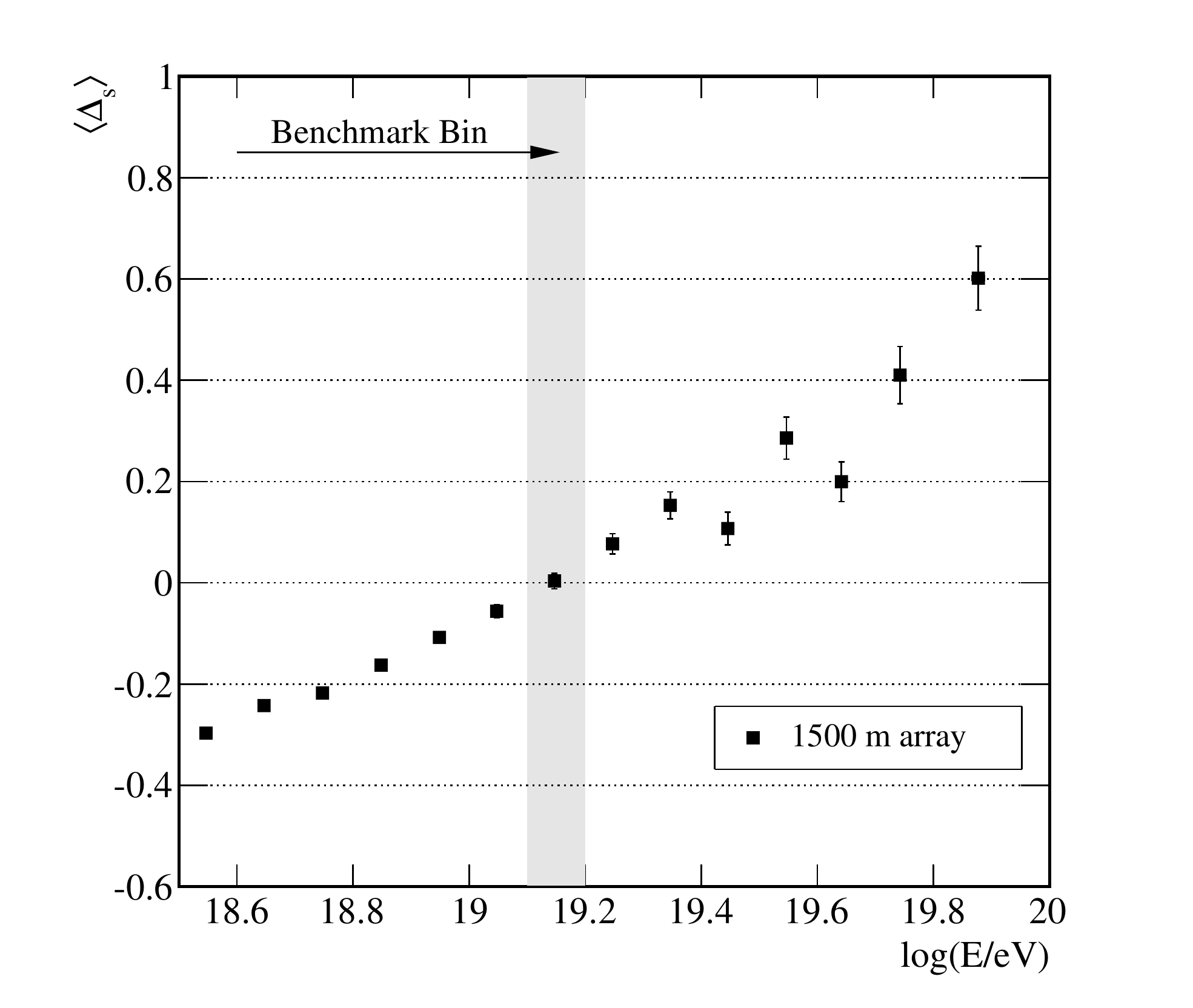}
\caption{Evolution of
$\langle$$\Delta_s$$\rangle$ as a function of energy for the two
surface arrays: 750 m (left), 1500 m (right). The grey bands show the
energy ranges where the benchmark functions were defined.}
\label{fig7}
\end{figure*}

We now describe the observed variation of $\langle$$\Delta_s$$\rangle$,
the mean of $\Delta_s$ for a set of events, as a function
of energy. The selection criteria for this analysis were presented in
Table \ref{t2}.
The variation of $\langle$$\Delta_s$$\rangle$ with energy for the two
arrays is shown in Fig. \ref{fig7}. Note that at the benchmark energies, indicated by
the vertical bands, $\langle$$\Delta_s$$\rangle$ is zero, as expected
by definition.

The results shown in Fig. \ref{fig7} were obtained using the whole data set. We produce
similar plots but this time splitting data in different bands of sec
$\theta$. This exercise gives results that are consistent with the ones
displayed in Fig. \ref{fig7}. Searches for anomalous behaviour of the largest, the
second largest and the smallest signals separately have also been made:
none was found.

To test the validity of hadronic models we can use
$\langle$$\Delta_s$$\rangle$. In previous works \cite{r4,r5,r6,r7} strong evidence has
been found that the models do not adequately describe the data and that
the problem lies with the predictions of the muon content of showers. As
the risetime is dominated by muons, $\langle$$\Delta_s$$\rangle$ is
expected to provide a further investigation of this problem that will be
useful because of the higher number of events and the extension to lower
energies.

Libraries of simulations for the QGSJETII-04 \cite{r19} and EPOS-LHC \cite{r20} models and
proton and iron primaries for zenith angles \textless{} 45\textsuperscript{o} and 17.5
\textless{} log (\textit{E}/eV) \textless{} 20 have been created. In
making comparisons with data it is necessary to choose which benchmarks
to adopt. For consistency in what follows we use the benchmarks
determined from data (section \ref{s4}). Different choices of benchmarks would
simply give shifts in the values of $\langle$$\Delta_s$$\rangle$, which
would be the same for each data set.

\begin{figure}
\centering
\includegraphics[width=\columnwidth]{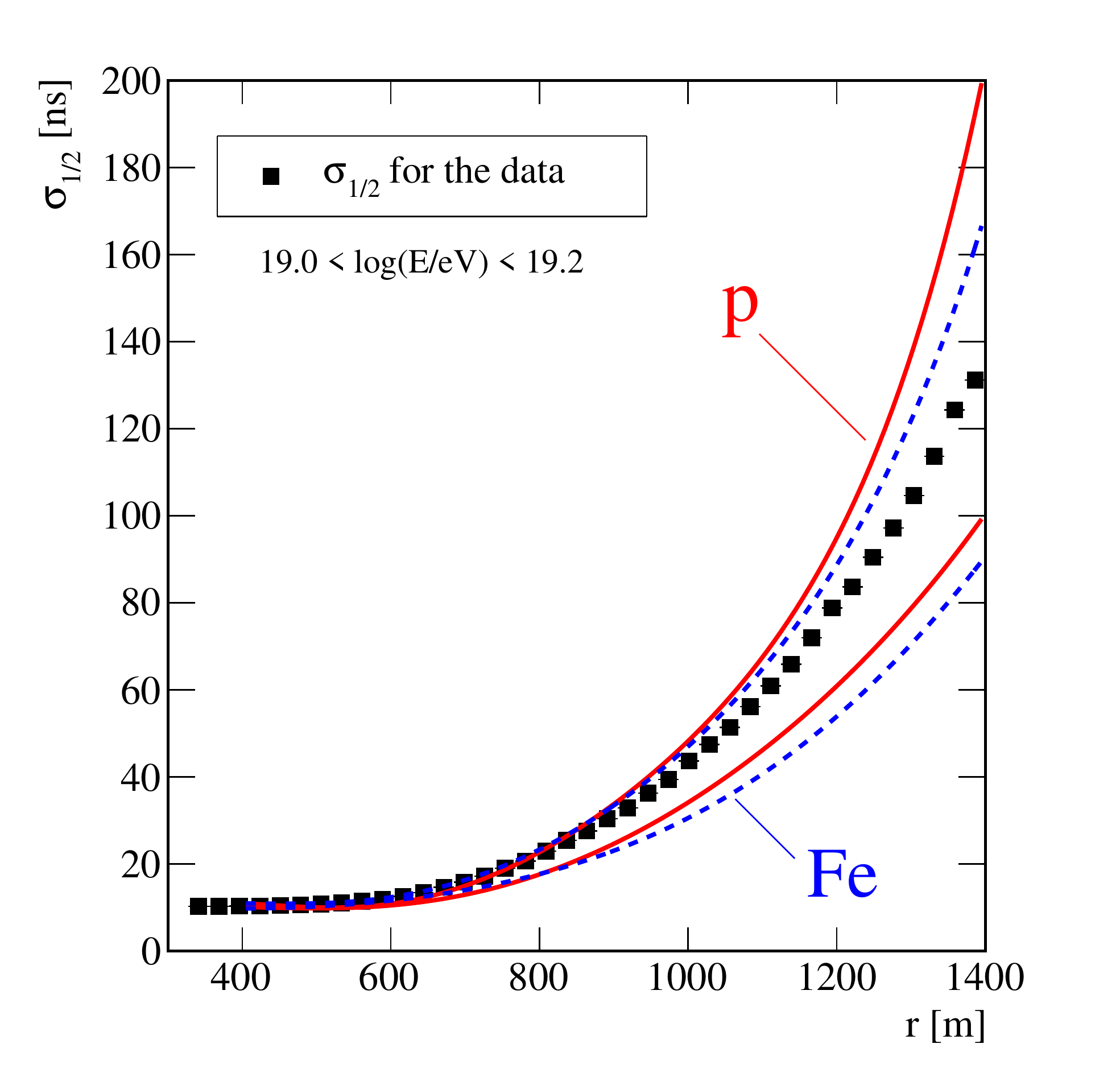}
\caption{Risetime uncertainties
estimated for protons (red lines), iron nuclei (blue dashed lines) and
data as a function of core distance. For a clear view, the uncertainties
corresponding to data are the average values. The uncertainties have
been evaluated for events with energies in the range 19.0 \textless{}
log (\textit{E}/eV) \textless{} 19.2. The regions bracketed by the lines
indicate the spread of the events simulated with QGSJETII-04 at a given
distance.}
\label{fig8}
\end{figure}

For this study, the uncertainties in the risetimes have been found from
simulations and adopting the `twins' approach described in section \ref{s3}.
The results are shown in Fig. \ref{fig8} where it is seen that the uncertainties from
the data are in good agreement with simulations using the QGSJETII-04
model at the benchmark energy.

\begin{figure*}
\centering
\includegraphics[width=\columnwidth]{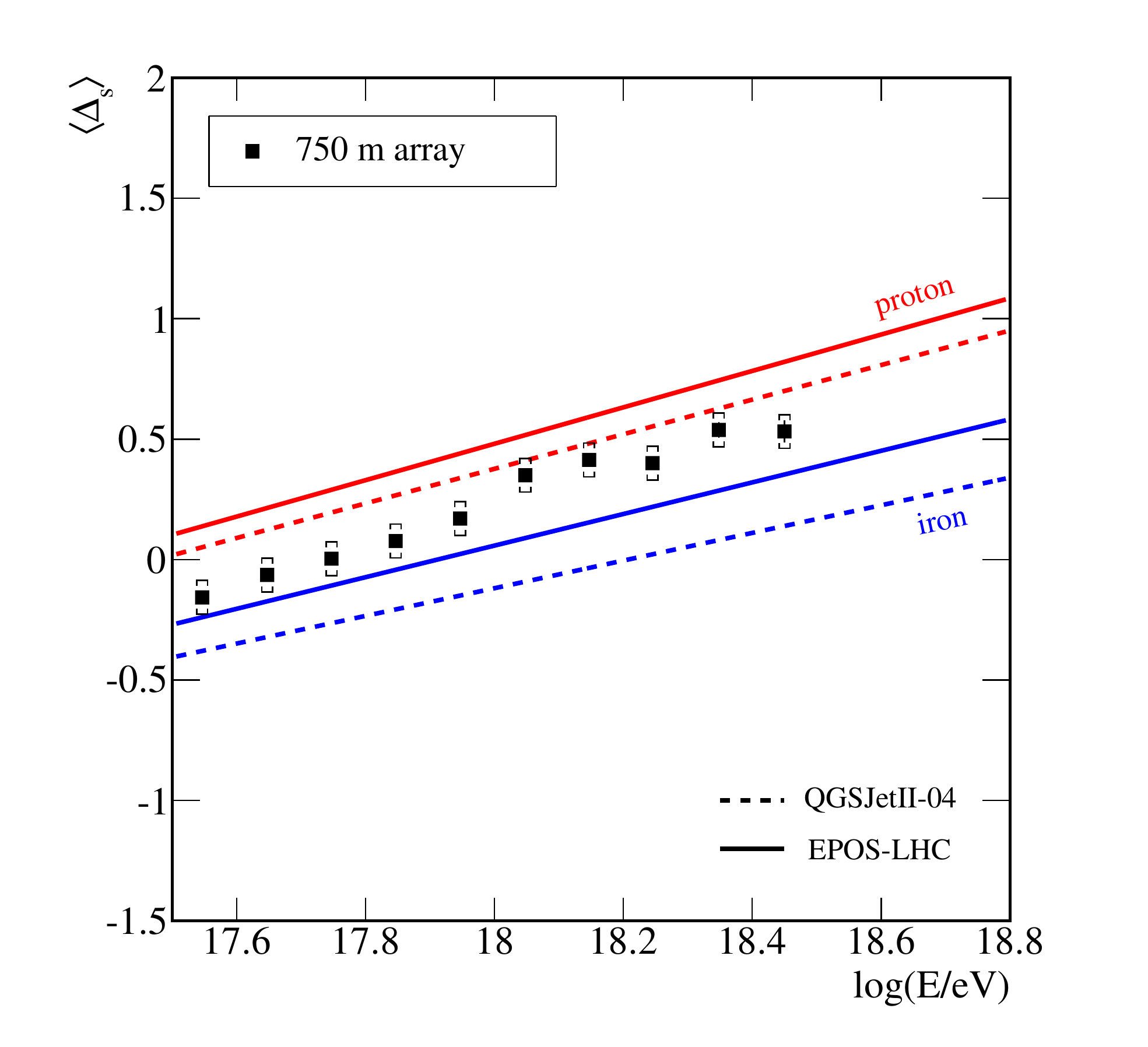}
\includegraphics[width=\columnwidth]{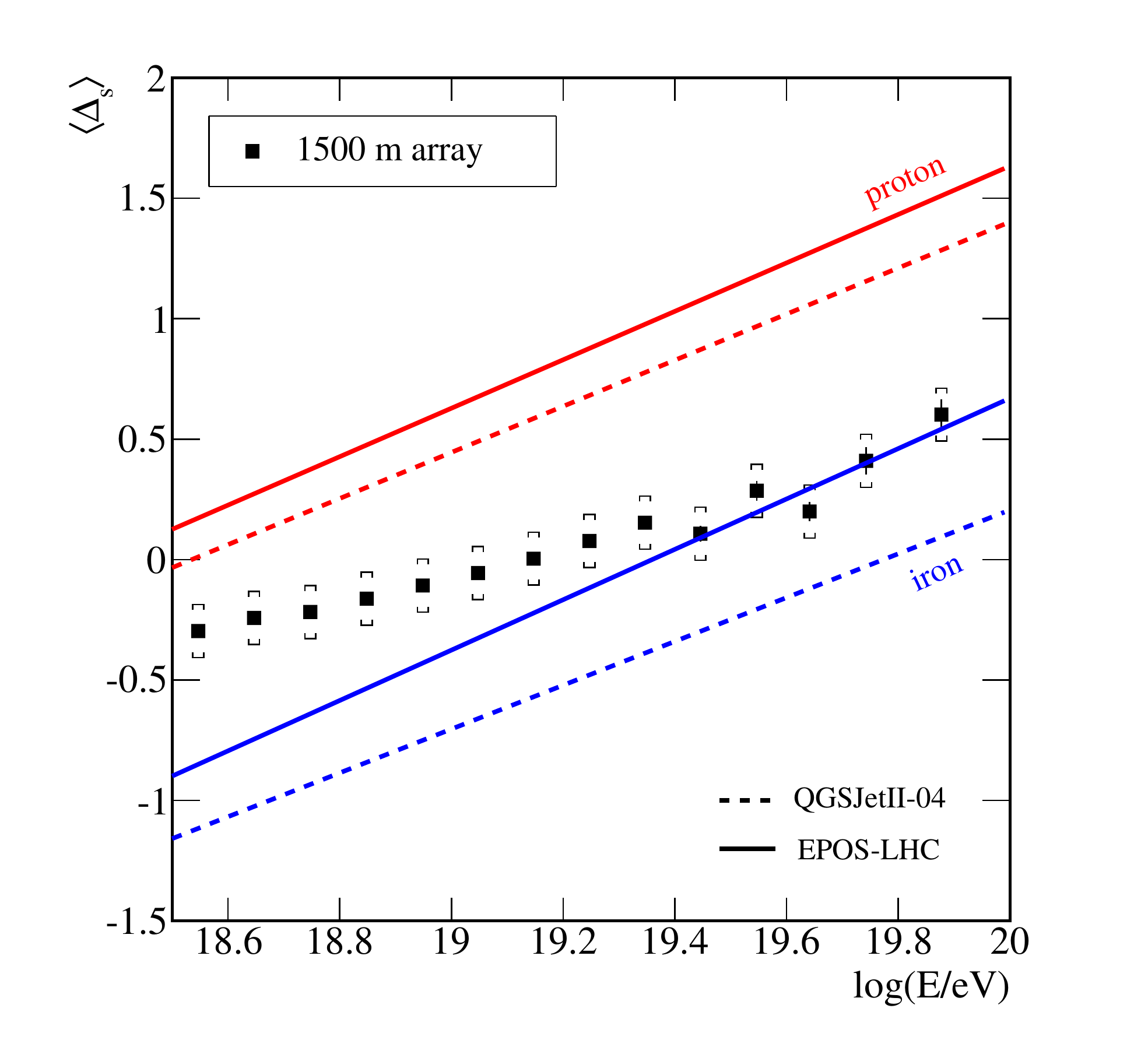}
\caption{$\langle$$\Delta_s$$\rangle$ as a function of the energy for the two
surface arrays. Brackets correspond to the systematic uncertainties.
Data are compared to the predictions obtained from simulations.}
\label{fig9}
\end{figure*}

A comparison of the evolution of $\langle$$\Delta_s$$\rangle$ with
energy from the data with those from models is shown in Fig. \ref{fig9}.

The main sources that contribute to the systematic uncertainty are: a
seasonal effect found when data are grouped according to the season of
the year. It amounts to 0.03 for the 1500 m data and it is due to the
variable conditions of pressure and temperature found in the atmosphere
through the year. The UTC time at which the data were recorded also
introduces a small uncertainty in our determination of
$\langle$$\Delta_s$$\rangle$. Splitting data into periods corresponding
to day and night, we obtain a value of this uncertainty of 0.01 for the
1500 m array data. Our observable also exhibits dependence with the
ageing effects of surface detectors. We take as a systematic uncertainty
the difference in $\langle$$\Delta_s$$\rangle$ found after grouping our
data into two samples, one running from 2004 to 2010 and the other one
from the years 2012 to 2014. For the 1500 m array, the difference
amounts to 0.04. A small dependence of $\langle$$\Delta_s$$\rangle$
with sec $\theta$ is taken as source of systematics, its value being
0.02. Finally the systematic uncertainty associated to the energy scale
($\pm$14\%) results in a systematic uncertainty on
$\langle$$\Delta_s$$\rangle$ that amounts to 0.1. Adding all these
contributions in quadrature, the overall systematic uncertainty in
$\langle$$\Delta_s$$\rangle$ is 0.11 for the 1500 m array. A similar
study for the 750 m array gives an overall systematic uncertainty in
$\langle$$\Delta_s$$\rangle$ of 0.07. According to simulations, this is
about 10\% of the separation between proton and iron nuclei. It is
evident, independent of which model is adopted, that the measurements
suggest an increase of the mean mass with energy above
$\sim$2.5 EeV \textit{if} the hadronic models are correct.

Assuming the superposition model is valid and since
$\langle$$\Delta_s$$\rangle$ is proportional to the logarithm of the
energy (Fig. \ref{fig7}), the mean value of the natural logarithm of \textit{A} (the
atomic weight of an element) can be found from the following equation

\begin{equation}
\langle \ln A \rangle = \ln 56 \frac{\langle \Delta_s \rangle_\text{p} - \langle \Delta_s \rangle_\text{data}}{\langle \Delta_s \rangle_\text{p} - \langle \Delta_s \rangle_\text{Fe}}
\end{equation}

\begin{figure}
\centering
\includegraphics[width=\columnwidth]{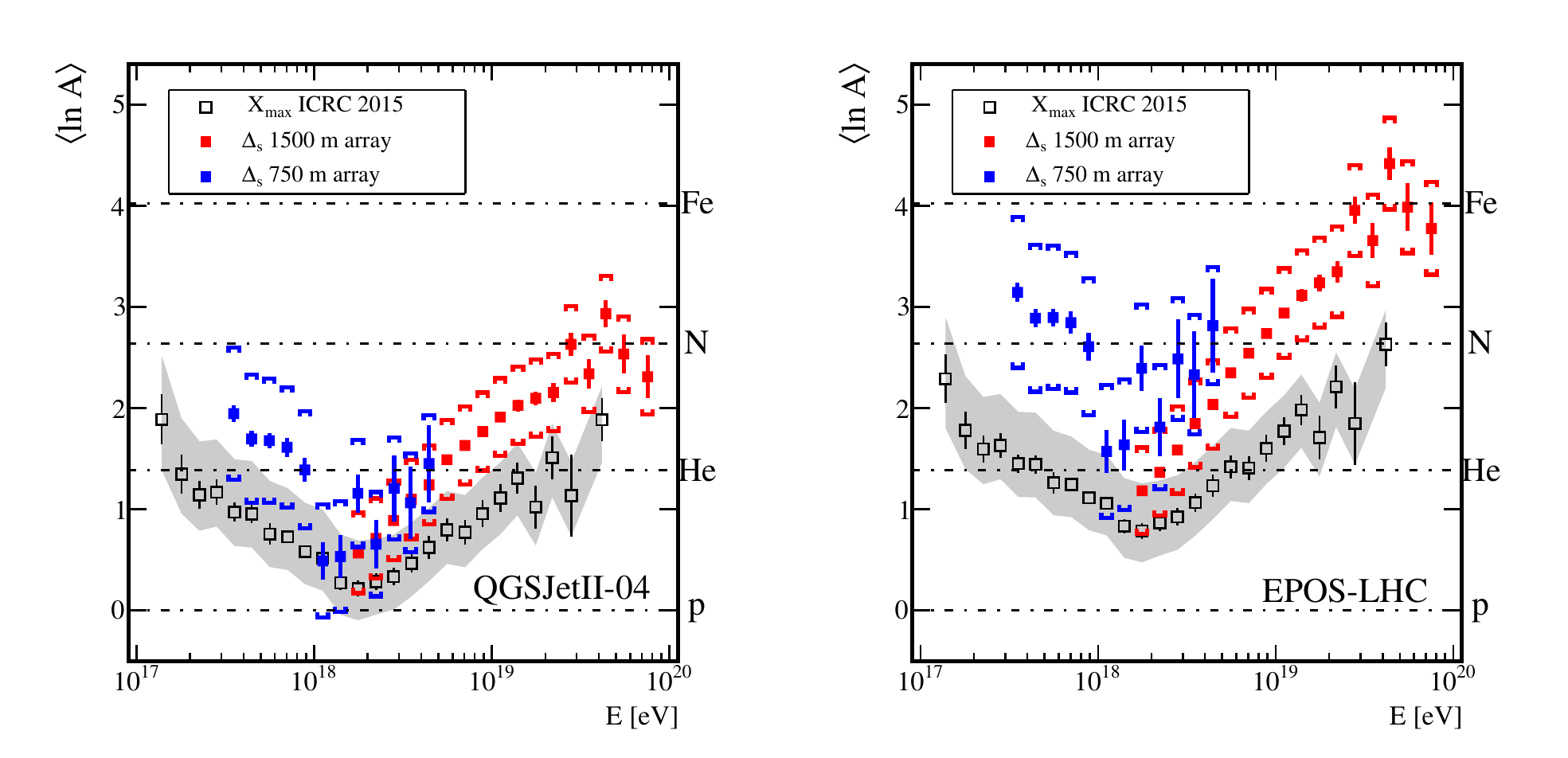}
\caption{$\langle$ln \textit{A}$\rangle$ as a
function of energy for the Delta method and for
\textit{X}\textsubscript{max} measurements done with the FD. QGSJetII-04
and EPOS-LHC have been used as the reference hadronic models.
Statistical uncertainties are shown as bars. Brackets and shaded areas
correspond to the systematic uncertainties associated to the
measurements done with the SD and FD data, respectively.}
\label{fig10}
\end{figure}

The results of this transformation for two models are shown in Fig. \ref{fig10}
and are
compared with the Auger measurements of \textit{X}\textsubscript{max} made
with the FD \cite{r21}. While the absolute values of $\langle$ln \textit{A}$\rangle$ for the Delta
method and the FD \textit{X}\textsubscript{max} differ from each other,
the trend in $\langle$ln \textit{A}$\rangle$ with energy is very similar. The observed
difference arises because of the inadequate description of the muon
component in the models used to get the $\langle$ln \textit{A}$\rangle$ values. Notice
that the electromagnetic cascade dominates the FD measurement whereas
the Delta method is of a parameter that is a mixture of muons and the
electromagnetic component. With substantially more events than in
previous studies, we observe that the inconsistency between data and
model predictions extends over a greater energy range than what was
probed in past works.

\begin{figure}
\centering
\includegraphics[width=\columnwidth]{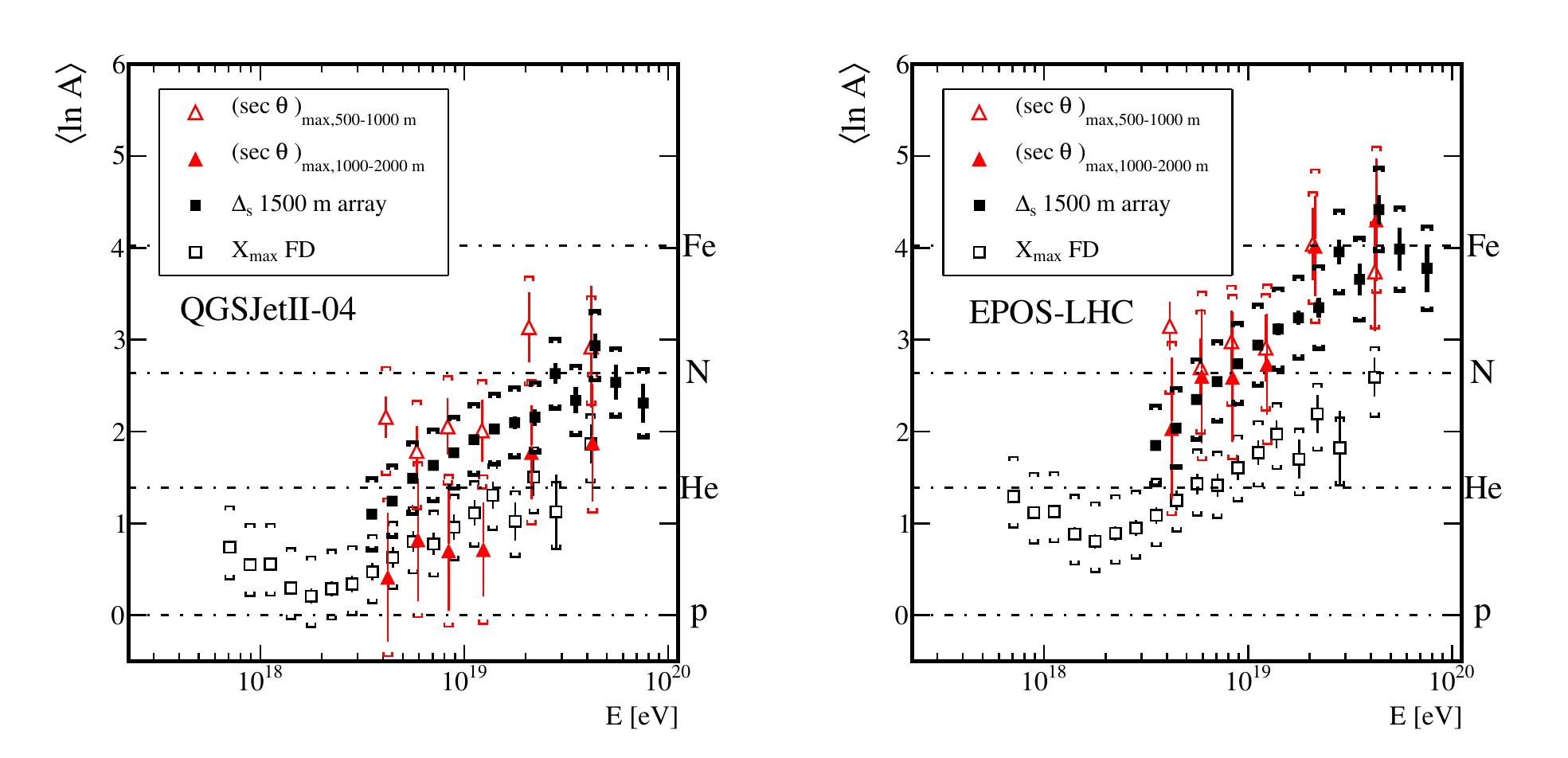}
\includegraphics[width=\columnwidth]{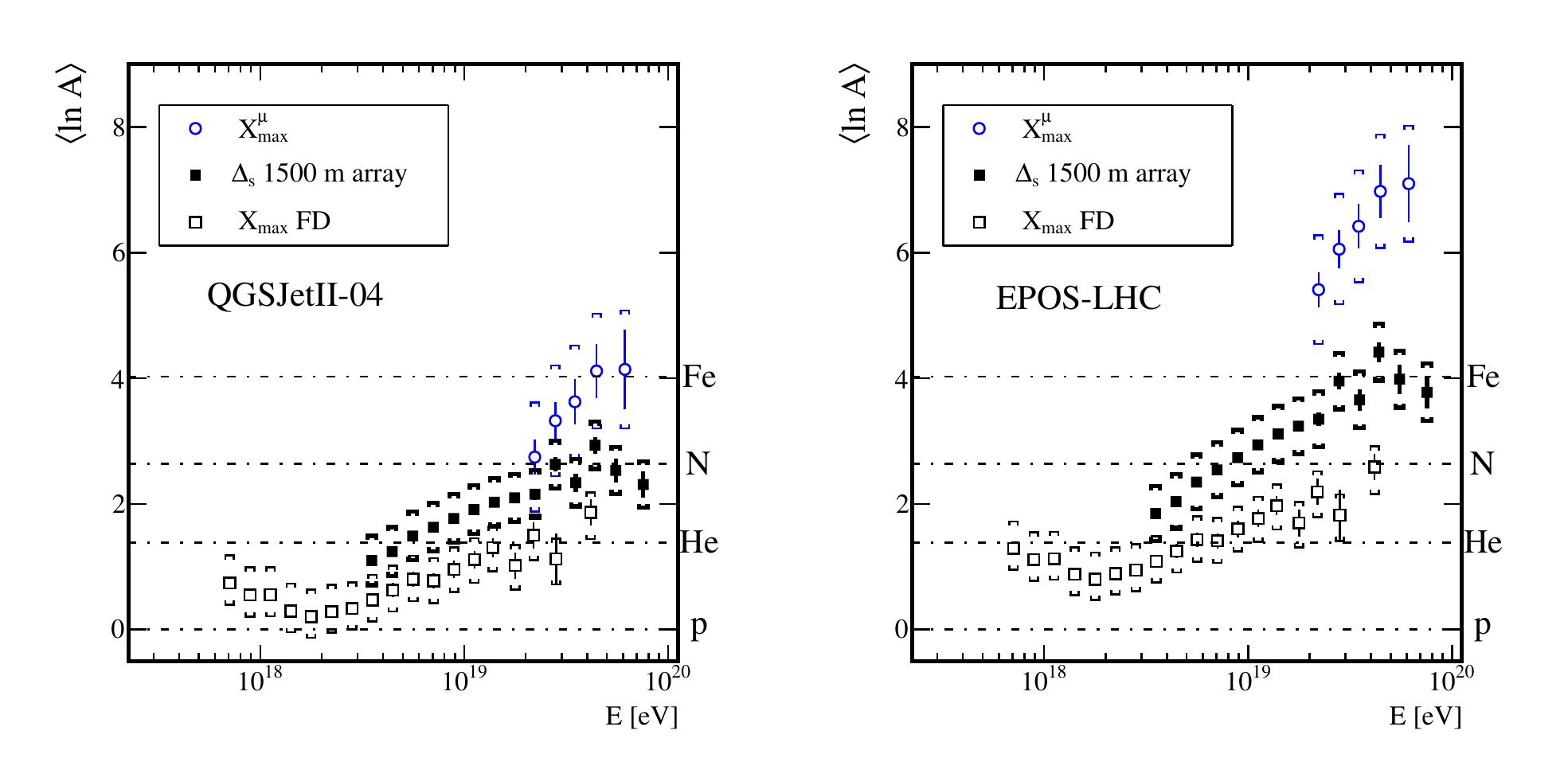}
\caption{$\langle$ln \textit{A}$\rangle$ as a
function of the energy for analyses using FD data and SD data from the
1500 m array. QGSJetII-04 and EPOS-LHC have been used as the reference
hadronic models. The results of the Delta method are compared with those
arising from the asymmetry analysis \cite{r6} (top panels) and
from the Muon Production Depth analysis \cite{r4} (bottom
panels). Brackets correspond to the systematic uncertainties.}
\label{fig11}
\end{figure}

In Fig. \ref{fig11}, the Delta results are also compared with the results of the
analysis made using the asymmetry method \cite{r6} and with those
from the study of the depth of muon production \cite{r4}

For EPOS-LHC the results from the asymmetry analysis, which is also
based on risetimes and consequently on signals which are a mixture of
the muon and the electromagnetic component, are in good agreement with
the Delta results, albeit within the rather large statistical
uncertainties. By contrast, the results from the MPD analysis, in which
only muons are studied, give much larger (and astrophysically
unexpected) values of $\langle$ln \textit{A}$\rangle$. This once more indicates that the
mechanisms of muon production in extensive air-showers are not properly
described in current hadronic models.

\section{Correlation of $\Delta_s$ with the Depth of Shower Maximum }
\label{s6}

We now address the correlation of $\Delta_s$ with the
depth of shower maximum, \textit{X}\textsubscript{max}. As remarked
earlier, we would not expect a 1:1 correlation between these parameters
because the muon/electromagnetic mix incident on the water-Cherenkov
detectors changes in a complex, but well-understood, manner with zenith
angle, energy and distance. An idea of the correlation to be expected
can be gained through Monte Carlo studies.

\begin{figure*}
\centering
\includegraphics[width=\columnwidth]{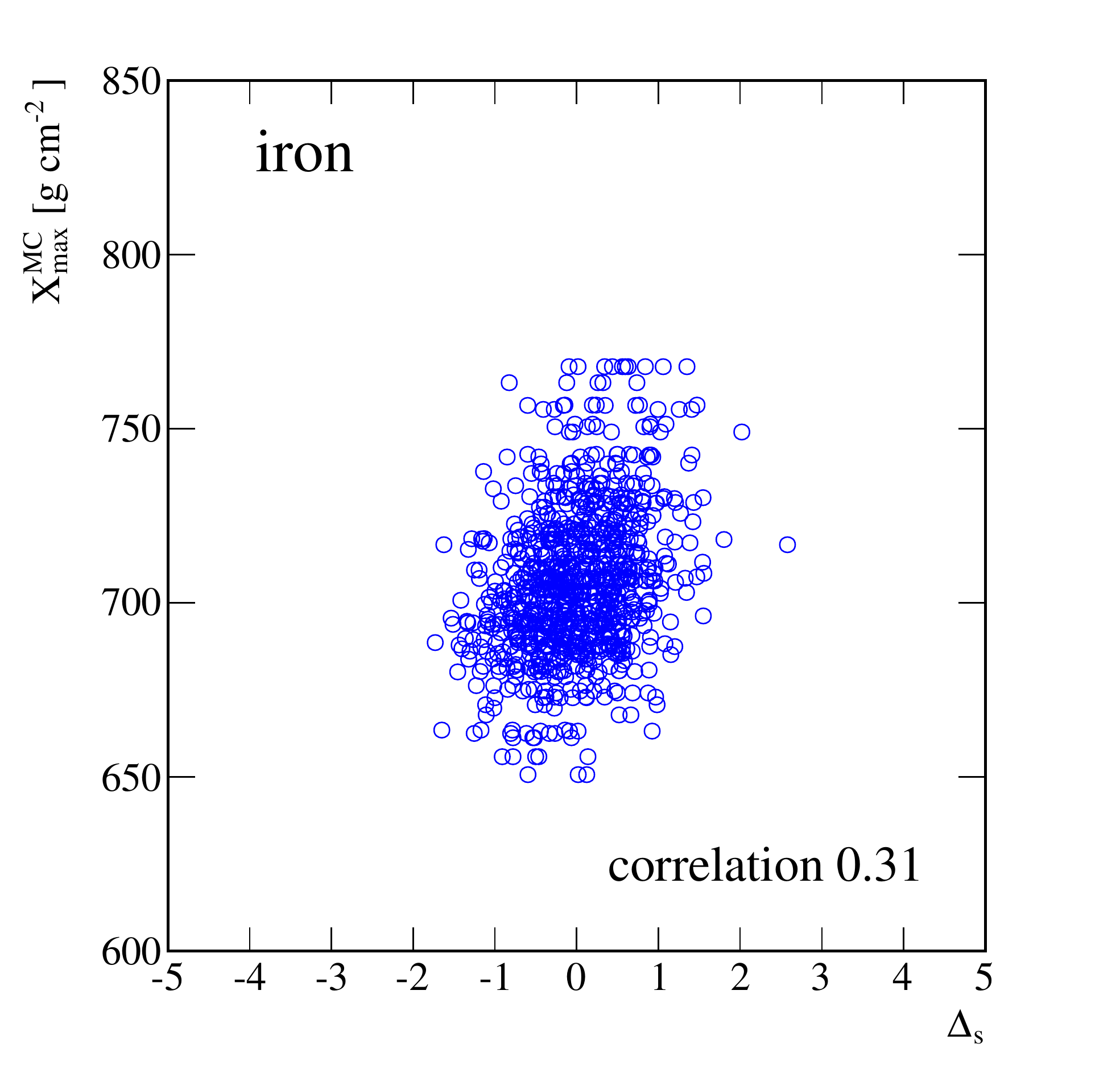}
\includegraphics[width=\columnwidth]{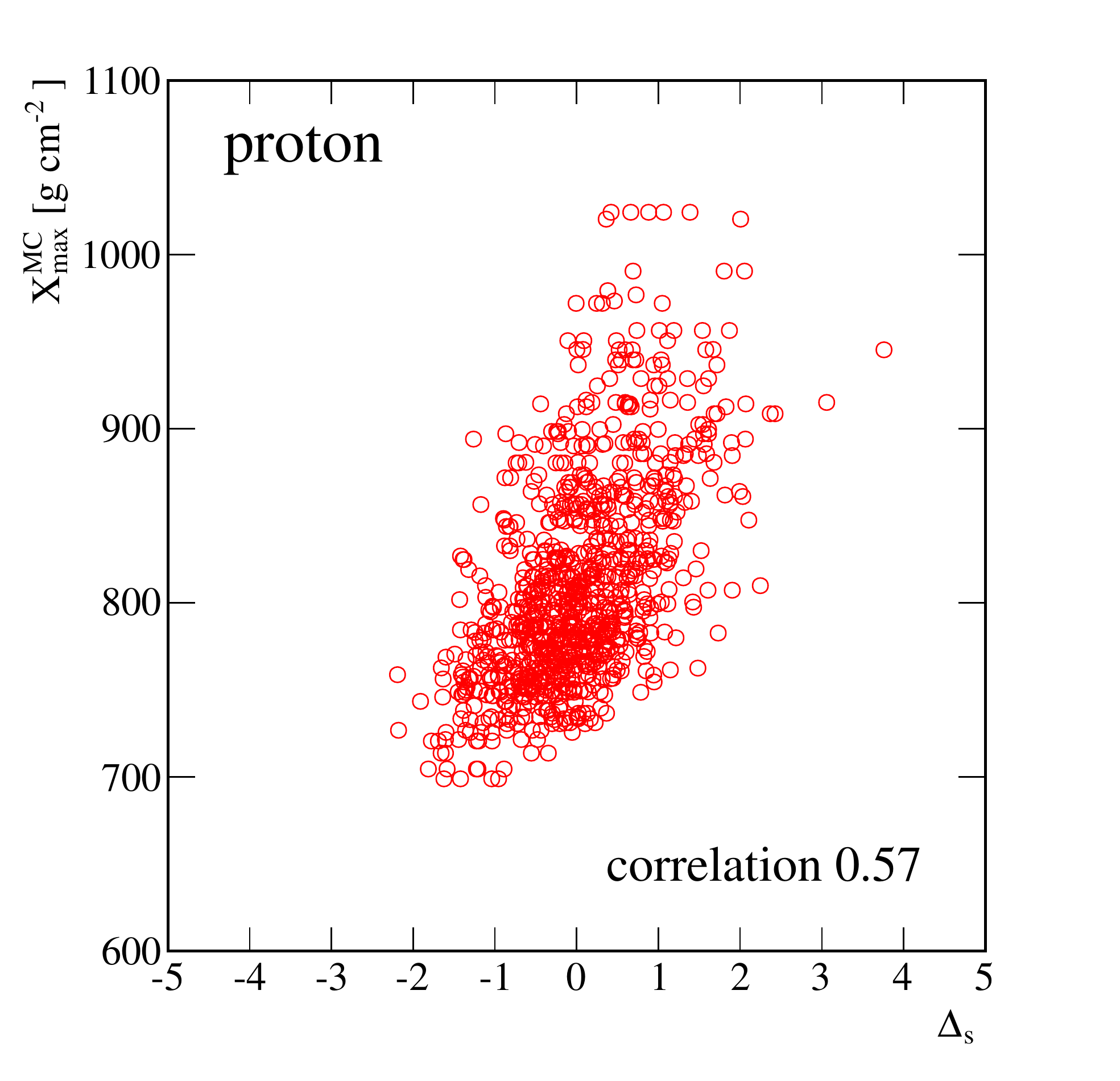}
\caption{Pearson's correlation of
$\Delta_s$ and the true values of
\textit{X}\textsubscript{max} for events simulated with QGSJETII-04 in the
energy range 19.1 \textless{} log (\textit{E}/eV) \textless{} 19.2. The
left panel corresponds to iron nuclei; the right panel shows the
correlation for protons. Values of $\Delta_s$ are
computed for three stations.}
\label{fig12}
\end{figure*}

Values of $\Delta_s$ and \textit{X}\textsubscript{max} have
been obtained from simulations of 1000 proton and 1000 iron nuclei
showers made using the QGSJETII-04 model for the benchmark bin of the
1500 m array. The results are shown for three stations in Fig. \ref{fig12}. The fact
that the Pearson's correlation is less strong for Fe-nuclei than for
protons, reflects the enhanced dominance of muons in showers initiated
by Fe-primaries. The simulations give an indication of what is to be
expected when the measurements of $\Delta_s$ are compared
with the \textit{X}\textsubscript{max} values in the hybrid events for
which the reconstruction of both observables is possible.

To exploit the correlation using data, and hence extend the energy range
and the statistical significance of the elongation rate determined with
the FD, it is necessary to create empirical correlations using events in
which both $\Delta_s$and \textit{X}\textsubscript{max}
have been measured in the same events. For this study we used the data
discussed in \cite{r1} for the 1500 m array for the events with energies
\textgreater{} 3 EeV and a similar set of data from the 750 m array
\cite{r21} for events of lower energy. The selection of events is
shown in Table \ref{t3}.

\begin{table*}
\caption{Set of cuts used to select
events simultaneously reconstructed by the fluorescence and surface
detectors. These events are used for calibration purposes. $\epsilon$
stands for the overall efficiency. HEAT data are obtained with a set of
three fluorescence detectors that point to the higher zenith angles
appropriate to the lower energies.}
\centering
\begin{tabular}{lrr||lrr}
\hline\hline
 \multicolumn{3}{c}{750 m array} & \multicolumn{3}{c}{1500 m array} \\
  \hline \hline 
\multicolumn{1}{l}{Quality cuts} & \multicolumn{1}{c}{Events} &
                                                                \multicolumn{1}{c||}{$\epsilon$
                                                                (\%)
                                                                }
                                                             & \multicolumn{1}{l}{Quality cuts} & \multicolumn{1}{c}{Events} & \multicolumn{1}{c}{$\epsilon$ (\%}) \\
HEAT data & 12 003 & 100.0 & FD data & 19 759 & 100.0\tabularnewline
FD \& SD recon & 2 461 & 20.5 & FD \& SD recon & 12 825 &
65.0\\
sec $\theta$ \textless{} 1.30 & 2 007 & 16.7 & sec $\theta$ \textless{}
1.45 & 9 625 & 49.0\\
6T5 trigger & 714 & 5.9 & 6T5 trigger & 7 361 & 37.0\tabularnewline
$\geq$ 3 selected stations & 660 & 5.5 & $\geq$ 3 selected stations & 4 025 &
20.0\\
log (\textit{E}/eV) $\geq$ 17.5 & \textbf{252} & \textbf{2.1} & log
(\textit{E}/eV) $\geq$ 18.5 & \textbf{885} & \textbf{4.5}\\
\hline\hline
\end{tabular}
\label{t3}
\end{table*}

\begin{figure*}
\centering
\includegraphics[width=\columnwidth]{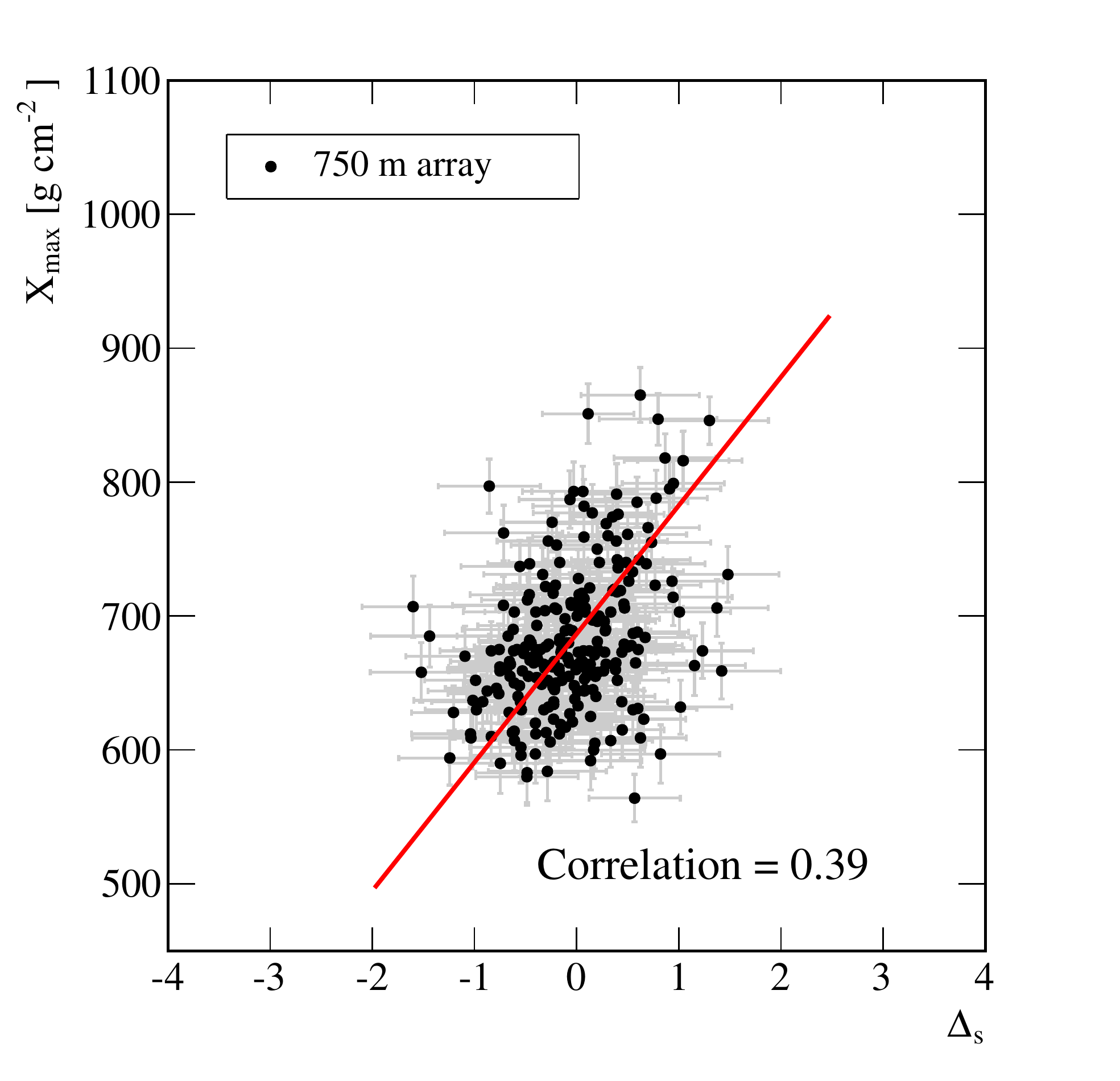}
\includegraphics[width=\columnwidth]{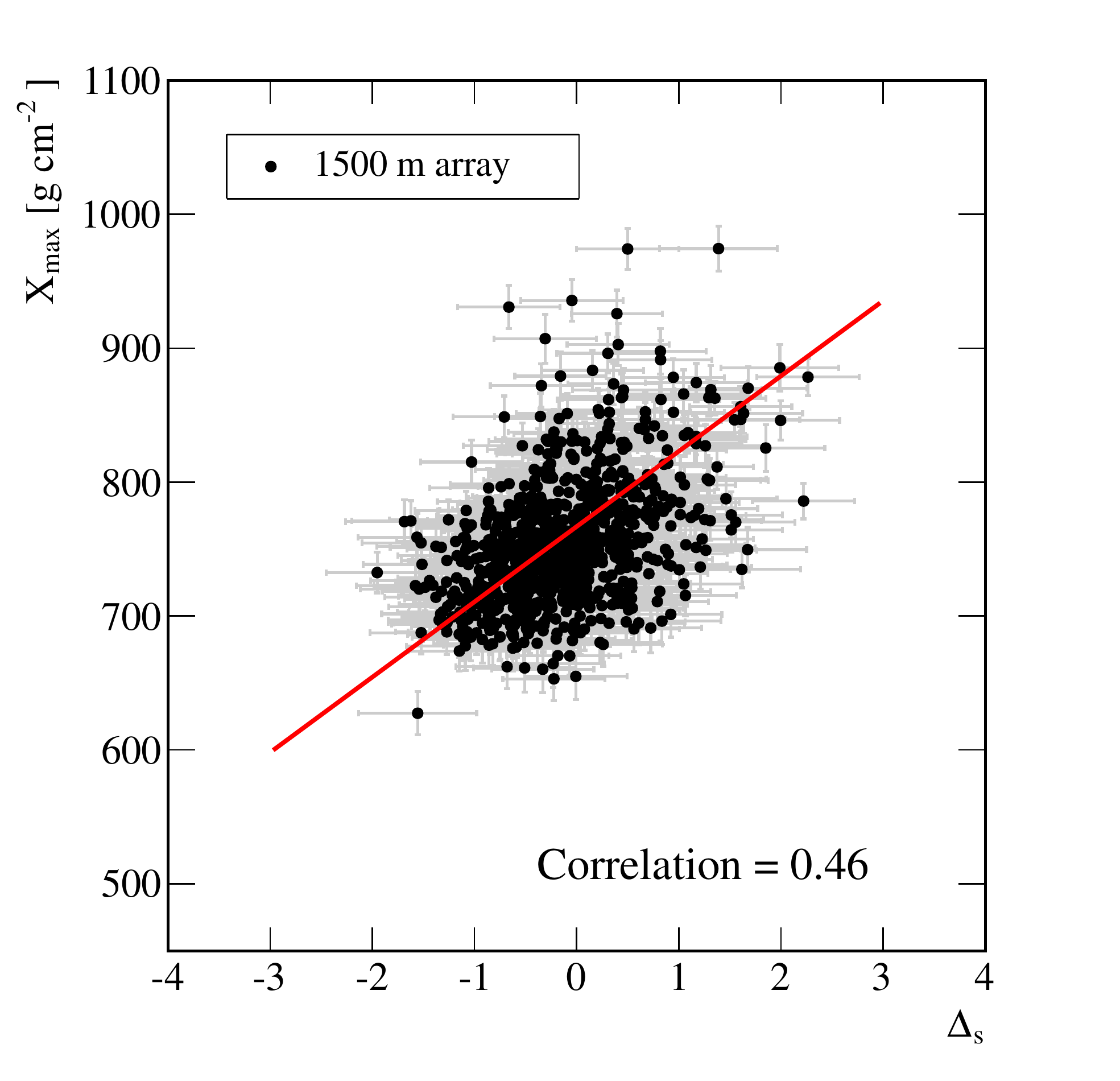}
\caption{ (Left) Correlation of
\textit{X}\textsubscript{max} and $\Delta_s$ for the 252
events from the 750 m array. (Right) Correlation of
\textit{X}\textsubscript{max} and $\Delta_s$for the 885
events of the 1500 m array.}
\label{fig13}
\end{figure*}

The $\Delta_s$and \textit{X}\textsubscript{max} of the
events selected for the purposes of calibration are shown for the two
arrays in Fig. \ref{fig13}. There are 252 and 885 events for the 750 m and 1500 m arrays
respectively available for calibration of which 161 have energies
\textgreater{}10 EeV. The small number for the 750 m array reflects the
shorter period of operation and the relatively small area (23.5
km\textsuperscript{2}) of the array. We have checked that the sample of
events selected is unbiased by comparing the elongation rate determined
from the full data set (from HEAT and standard fluorescence telescopes)
with that found from the 252 and 885 events alone.

\begin{table}
\caption{Coefficients obtained from
the calibration of $\Delta_s$ and
\textit{X}\textsubscript{max}.}
\centering
\begin{tabular}{lcc}
\hline\hline
& 750 m array & 1500 m array\\\hline\hline
Calibration parameters & Value (g cm$^{-2}$) & Value (g
cm $^{-2}$)\\\hline
\textit{a} & 636 $\pm$ 20 & 699 $\pm$ 12\\
\textit{b} & 96 $\pm$ 10 & 56 $\pm$ 3\\
\textit{c} & 2.9 $\pm$ 1.2 & 3.6 $\pm$ 0.7\\
\hline\hline
\end{tabular}
\label{t4}
\end{table}

For the calibration we fit functions of the form
\begin{equation}
X_\text{max} = a + b \Delta_s + c \log (E/\text{eV}) 
\end{equation}
to the two data sets. The term `\textit{b}' is dominant in the fit. The
term `\textit{c}' is included to accommodate the energy dependence of both
variables. A fit including a quadratic term in log (\textit{E}/eV) does not
modify our results. The uncertainties in \textit{X}\textsubscript{max} are
taken from \cite{r1}. We have used the maximum likelihood method to make the fits
which give the coefficients listed in Table \ref{t4}. The three coefficients are not
independent. Their Pearson's correlations are
$\rho_{ab}$=-0.2, $\rho_{ac}$=-0.97 and
$\rho_{bc}$=0.34. These correlations are taken into
account when evaluating the systematic uncertainty associated with the
calibration procedure.

We have also evaluated the systematic uncertainties associated with the
measurements of \textit{X}\textsubscript{max} deduced from the surface
detectors. These include the seasonal, diurnal, ageing and $\theta$
dependence already discussed for $\langle$$\Delta_s$$\rangle$ in
section \ref{s5} that \textit{X}\textsubscript{max} propagate to our measurement.
Now two further sources of systematic arise. One is related to the
uncertainty in the calibration parameters. We have propagated this
uncertainty taking into account the correlation of the parameters
\textit{a}, \textit{b} and \textit{c}. For the 1500 m array, the differences
in \textit{X}\textsubscript{max} span from 3 g cm\textsuperscript{-2} at
the lowest energies to 5 g cm\textsuperscript{-2} at the upper end of
the energy spectrum. We quote conservatively as a systematic uncertainty
the largest value found and consider it constant for the whole energy
range. A similar procedure for the 750 m array data results on a
systematic uncertainty of 10 g cm\textsuperscript{-2}. The systematic
uncertainty obtained in the measurement of \textit{X}\textsubscript{max}
with the FD detector propagates directly into the values obtained with
the SD data. In \cite{r1} the systematic uncertainty is given as
a function of the energy. In this analysis, the average of those values
is quoted a systematic uncertainty that is constant with energy. The
values are shown for each effect and for each array in Table \ref{t5}.

The systematic uncertainties have been added in quadrature to give 14
and 11 g cm\textsuperscript{-2} for the 750 and 1500 m arrays
respectively.

\begin{figure*}
\centering
\includegraphics[width=\columnwidth]{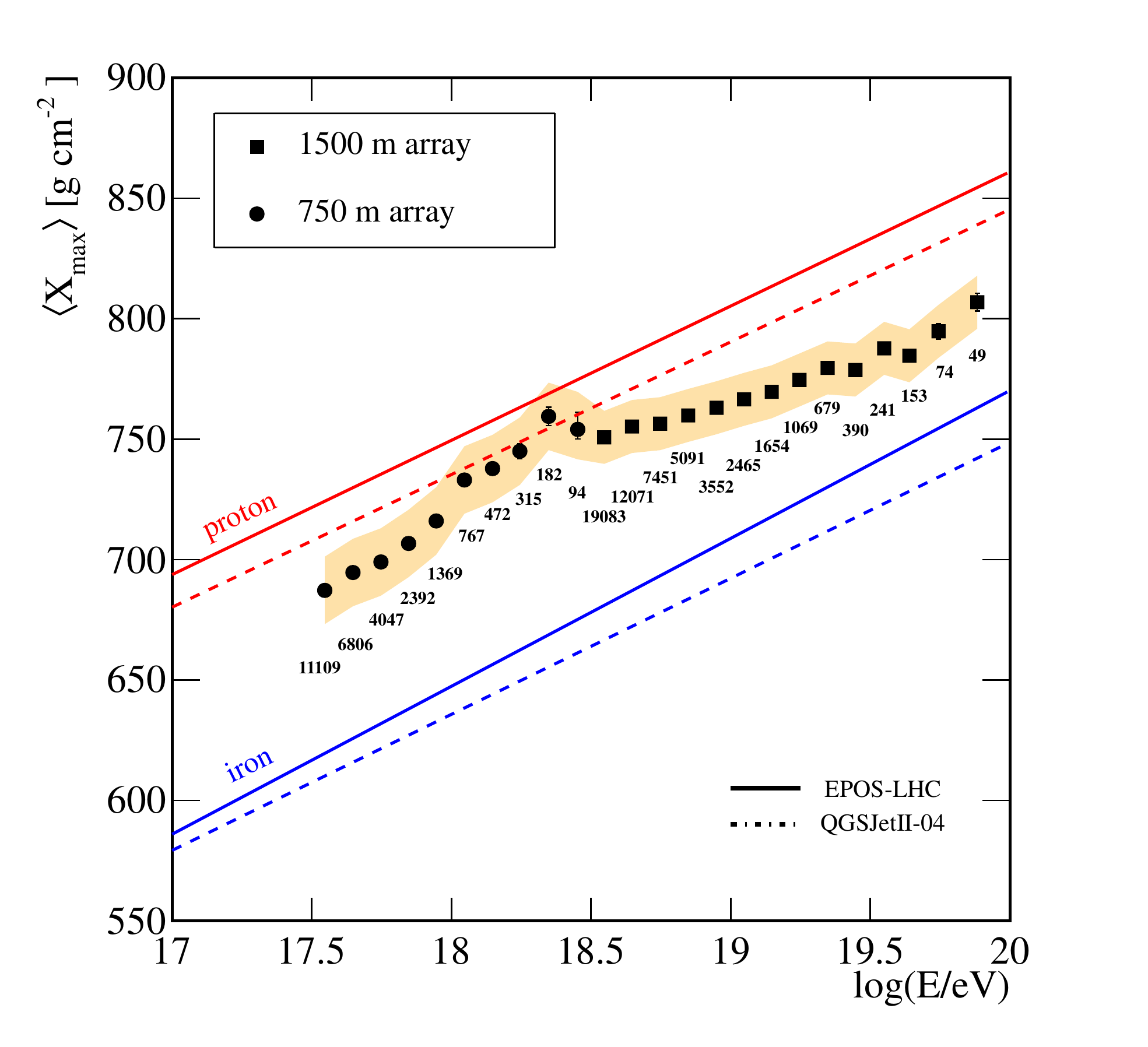}
\caption{Mean values of the
\textit{X}\textsubscript{max} distributions obtained with the data of the
750 m and 1500 m surface arrays as a function of the energy. The shaded
area indicates the systematic uncertainties. Data are compared to the
predictions from simulations of protons and iron nuclei for two
different hadronic models. The number of selected events in each energy
bin is indicated.}
\label{fig14}
\end{figure*}

The values of \textit{X}\textsubscript{max} found from this analysis are
shown as a function of energy in Fig. \ref{fig14}. The resolution in the measurement of
\textit{X}\textsubscript{max} with the surface detector data is 45 g
cm\textsuperscript{-2}.

\begin{table*}
\caption{Breakdown of the systematic
uncertainties of \textit{X}\textsubscript{max} for the 750 m and 1500
m arrays. The systematic uncertainty obtained in the measurement of \textit{X}\textsubscript{max}
with the FD and HEAT detectors propagates directly into the values obtained 
with the SD data. The rest of systematic uncertainties quoted in this table are 
intrinsic to the Delta method.}
\centering
\begin{tabular}{lc||lc}\hline\hline
\multicolumn{2}{c}{750 m array} & \multicolumn{2}{c}{1500 m array}\\\hline\hline
\begin{minipage}[t]{0.24\columnwidth}\raggedright\strut
Source\strut
\end{minipage} & \begin{minipage}[t]{0.24\columnwidth}\raggedright\strut
Systematic uncertainty

(g cm\textsuperscript{-2})\strut
\end{minipage} & \begin{minipage}[t]{0.24\columnwidth}\raggedright\strut
Source\strut
\end{minipage} & \begin{minipage}[t]{0.24\columnwidth}\raggedright\strut
Systematic uncertainty

(g cm\textsuperscript{-2})\strut
\end{minipage}\tabularnewline
Uncertainty on calibration & 10.0 & Uncertainty on calibration &
5.0\\
Seasonal effect & 2.0 & Seasonal effect & 2.0\\
Diurnal dependence & 1.0 & Diurnal dependence & 1.0\\
Ageing & 3.0 & Ageing & 3.0\\
HEAT systematic uncertainty & 8.5 & FD systematic uncertainty &
8.5\\
Angular dependence & \textless{}1.0 & Angular dependence &
1.5\\
\textbf{Total} & \textbf{14.0} & \textbf{Total} &
\textbf{11.0}\\\hline\hline
\end{tabular}
\label{t5}
\end{table*}

\begin{figure*}
\centering
\includegraphics[width=\columnwidth]{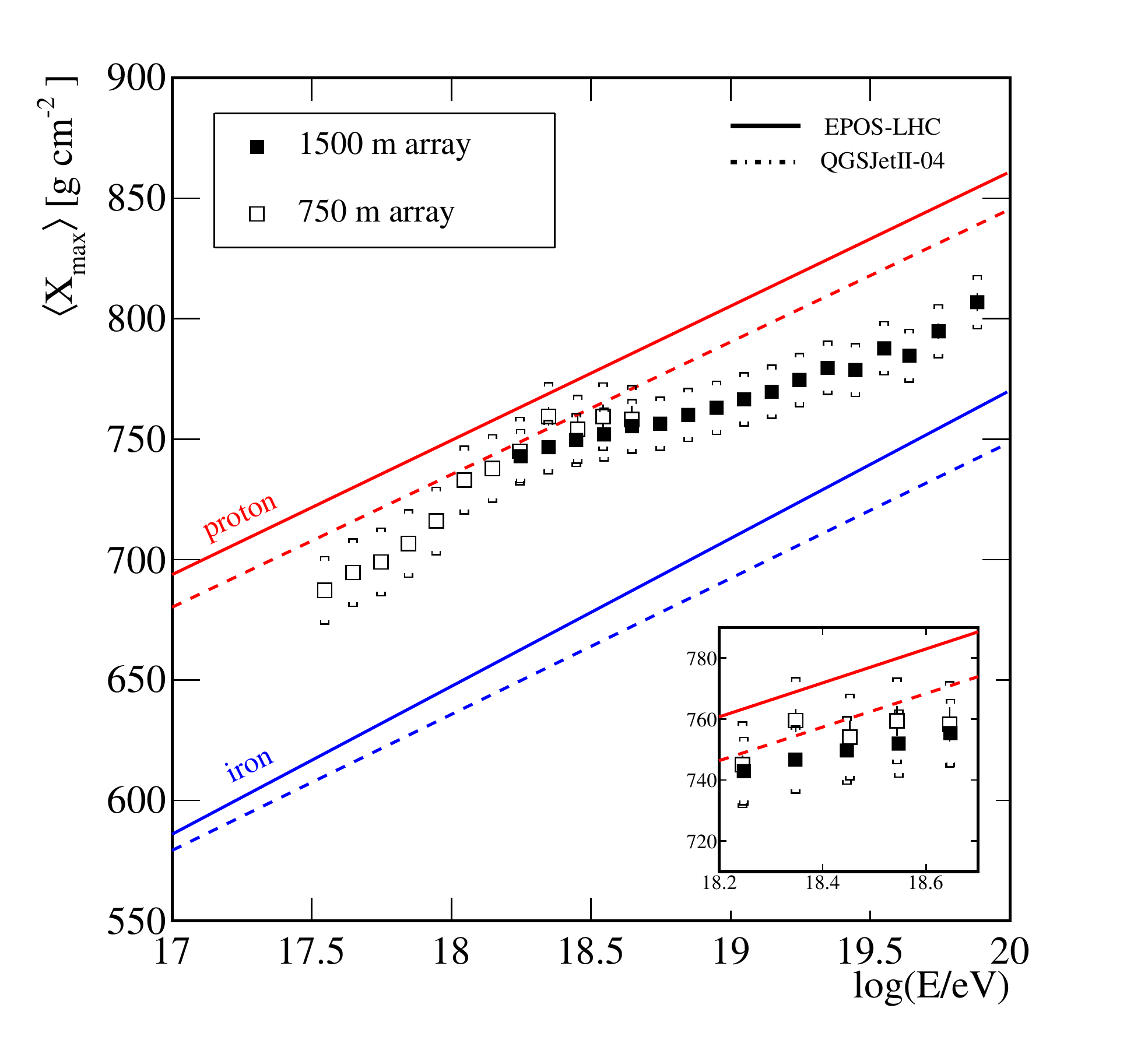}
\caption{Same as Fig. \ref{fig14} including
additional \textit{X}\textsubscript{max} measurements from the surface
detectors above and below 3 EeV. (Inset) Three energy bins have been
included below 3 EeV using the data of the 1500 m array and two
measurements added above 3 EeV use the data of the 750 m array. There is
good agreement between measurements in the overlap region. The brackets
correspond to systematic uncertainties.}
\label{fig15}
\end{figure*}

In Fig. \ref{fig15} measurements in the region of overlap between the two arrays
are shown. The agreement is satisfactory.

\begin{figure*}
\centering
\includegraphics[width=\columnwidth]{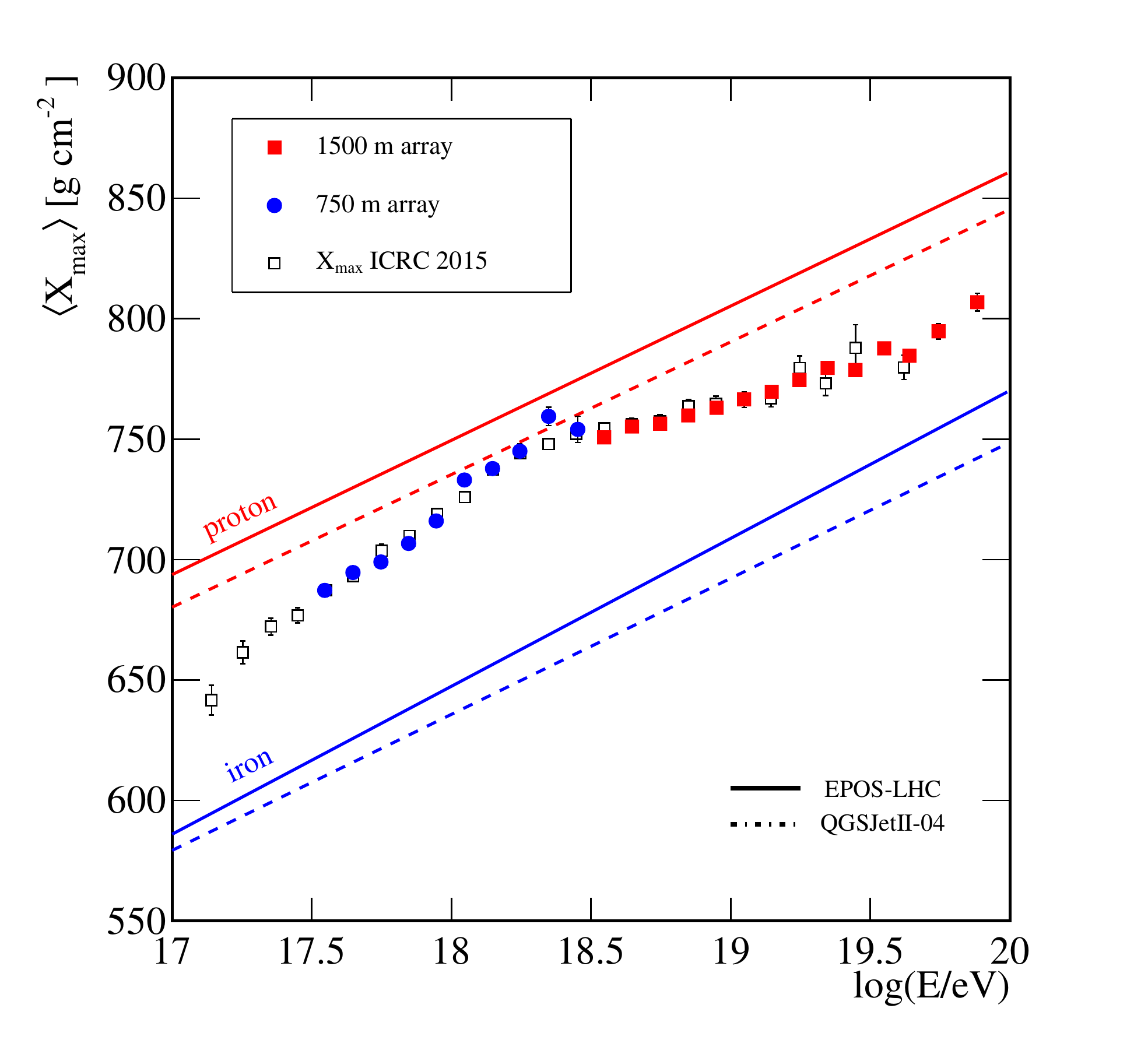}
\caption{Evolution of
$\langle$\textit{X}\textsubscript{max}$\rangle$ as a function of energy. The figure
compares the mean values of the \textit{X}\textsubscript{max}
distributions measured by the fluorescence and surface detectors of the
Pierre Auger Observatory. In most cases the uncertainties are smaller
than the size of the symbols.}
\label{fig16}
\end{figure*}

In Fig. \ref{fig16} the data of Fig. \ref{fig14} are compared with measurements made with
the fluorescence detectors {[}\textbf{21}{]}.

The agreement is good: the results from the surface detector alone are
statistically stronger and extend to higher energies.

\subsection{Interpretation of the measurements in terms of average mass}

\begin{figure}
\centering
\includegraphics[width=\columnwidth]{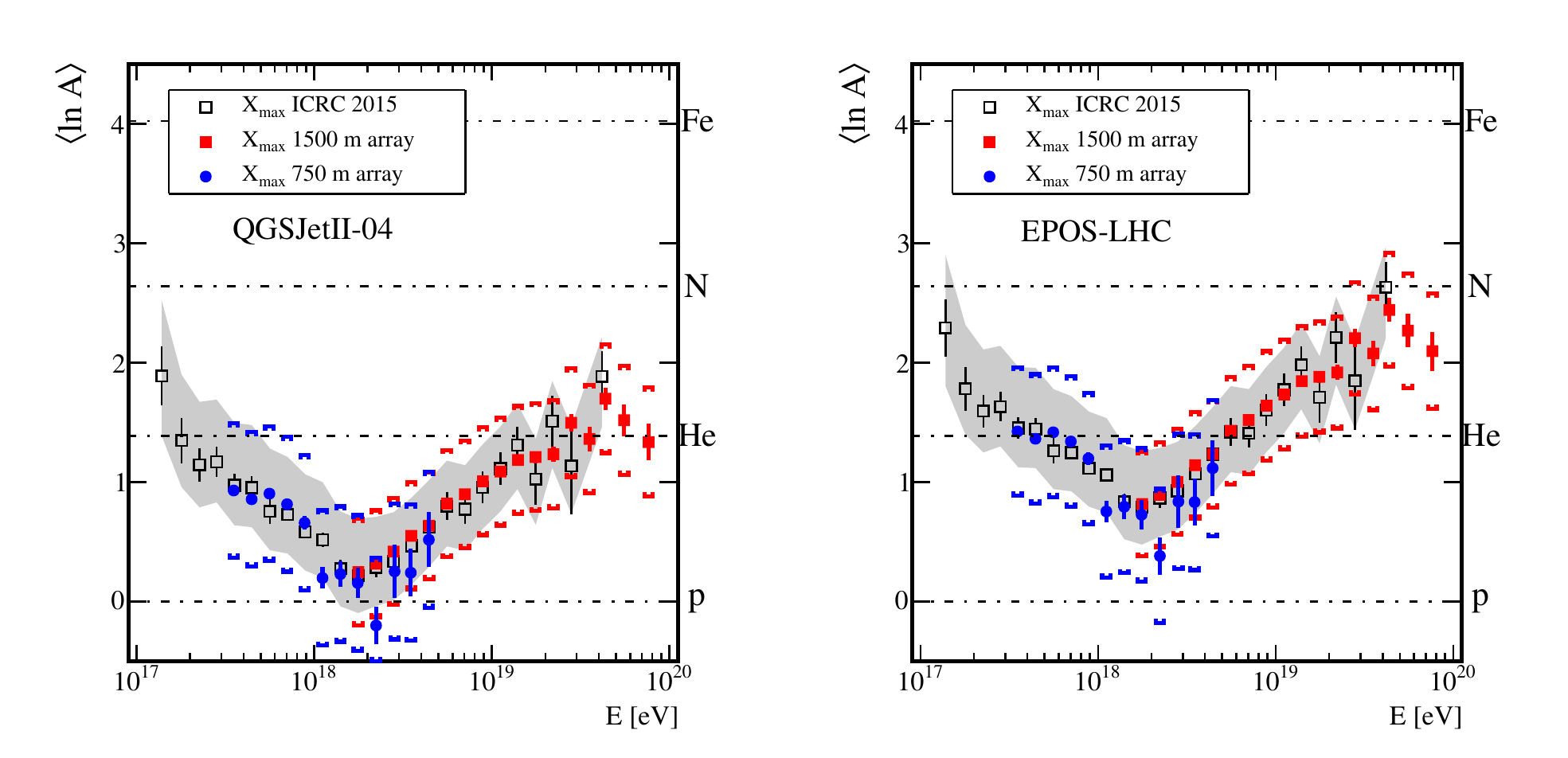}
\caption{$\langle$ln \textit{A}$\rangle$ as a
function of the energy. QGSJetII-04 and EPOS-LHC have been used as the
reference hadronic models. The results of the Delta method are compared
with those based on \textit{X}\textsubscript{max} measurements done with
the FD \cite{r21}. Brackets and shaded areas correspond to the
systematic uncertainties; bars correspond to uncertainties of
statistical nature.}
\label{fig17}
\end{figure}

A comparison with hadronic models allows the expression of the average
depth of shower maxima in terms of the natural logarithm of the atomic
mass $\langle$ln \textit{A}$\rangle$, following the procedure discussed in section \ref{s5}.
The evolution of $\langle$ln \textit{A}$\rangle$ as a function of
energy is shown in Fig. \ref{fig17}.
In the energy range where the FD and SD measurements coincide, the
agreement is good. For both hadronic models the evolution of $\langle$ln
\textit{A}$\rangle$ with energy is similar. However the EPOS-LHC model suggests a
heavier average composition. SD measurements have been used to confirm,
with a larger data set, what has already been observed with FD
measurements, namely that the primary flux of particles is predominantly
composed of light particles at around 2 EeV and that the average mass
increases up to $\sim$40 EeV. Above this energy, the SD
measurements can be used to draw inferences about mass composition with
good statistical power. The last two bins indicate a possible change in
the dependence of \textit{X}\textsubscript{max} with energy above 50 EeV,
with the final point lying $\sim$3 sigma above the elongation
rate fitted to data above 3 EeV. It is, therefore, possible that the
increase of the primary mass with energy is slowing at the highest
energies but we need to reduce statistical and systematic uncertainties
further before strong conclusions can be drawn. AugerPrime, the upgrade
of the surface-detector array of the Pierre Auger Observatory \cite{r22}, will
significantly improve our capability to elucidate mass composition on an
event-by-event basis in the energy range of the flux suppression.

\section{Summary and Conclusions}
\label{s7}

We have described a new method for extracting relevant information from
the time profiles of the signals from the water-Cherenkov detectors of
the Pierre Auger Observatory. With it, we have been able to obtain
information on the evolution of the mean depth of shower maximum with
energy over a larger energy range than has been studied previously using
over 81,000 events of which 123 are of energy \textgreater{}50 EeV. We
have also been able to expand the discussions of the mismatch between
data and predictions from models based on extrapolations of hadronic
interactions from LHC energies. Specifically we have reported the
following:

\begin{enumerate}
\def\labelenumi{\arabic{enumi}.}
\item
  The comparison of the risetime data with fluorescence measurements
  reinforces the conclusions reported previously \cite{r4,r5,r6,r7} that the modelling of
  showers provides an inadequate description of air-shower data. The
  deductions are made over a larger energy range and with smaller
  statistical uncertainties than hitherto (Fig. \ref{fig10} and Fig. \ref{fig11}).
\item
  The depth of shower maximum has been measured from 0.3 EeV to 100 EeV
  using data of the Surface Detector (Fig. \ref{fig14}).
\item
  Data from the 750 m array of the Observatory have been used to derive
  mass information for the first time.
\item
  The mean measurements of \textit{X}\textsubscript{max} have been
  compared with predictions from the EPOS-LHC and QGSJetII04 models and
  estimates of $\langle$ln \textit{A}$\rangle$ extracted (Fig. \ref{fig17}). While the EPOS-LHC model
  leads to larger values of $\langle$ln \textit{A}$\rangle$ than are found with the
  other model, both show the general trend of the mean mass becoming
  smaller as the energy increases up to $\sim$2 EeV, after
  which it rises slowly with energy up to about 50 EeV where this rise
  seems to stop.
\end{enumerate}


\section*{Acknowledgments}

\begin{sloppypar}
The successful installation, commissioning, and operation of the Pierre Auger Observatory would not have been possible without the strong commitment and effort from the technical and administrative staff in Malarg\"ue. We are very grateful to the following agencies and organizations for financial support:
\end{sloppypar}

\begin{sloppypar}
Argentina -- Comisi\'on Nacional de Energ\'\i{}a At\'omica; Agencia Nacional de Promoci\'on Cient\'\i{}fica y Tecnol\'ogica (ANPCyT); Consejo Nacional de Investigaciones Cient\'\i{}ficas y T\'ecnicas (CONICET); Gobierno de la Provincia de Mendoza; Municipalidad de Malarg\"ue; NDM Holdings and Valle Las Le\~nas; in gratitude for their continuing cooperation over land access; Australia -- the Australian Research Council; Brazil -- Conselho Nacional de Desenvolvimento Cient\'\i{}fico e Tecnol\'ogico (CNPq); Financiadora de Estudos e Projetos (FINEP); Funda\c{c}\~ao de Amparo \`a Pesquisa do Estado de Rio de Janeiro (FAPERJ); S\~ao Paulo Research Foundation (FAPESP) Grants No.\ 2010/07359-6 and No.\ 1999/05404-3; Minist\'erio de Ci\^encia e Tecnologia (MCT); Czech Republic -- Grant No.\ MSMT CR LG15014, LO1305, LM2015038 and CZ.02.1.01/0.0/0.0/16\_013/0001402; France -- Centre de Calcul IN2P3/CNRS; Centre National de la Recherche Scientifique (CNRS); Conseil R\'egional Ile-de-France; D\'epartement Physique Nucl\'eaire et Corpusculaire (PNC-IN2P3/CNRS); D\'epartement Sciences de l'Univers (SDU-INSU/CNRS); Institut Lagrange de Paris (ILP) Grant No.\ LABEX ANR-10-LABX-63 within the Investissements d'Avenir Programme Grant No.\ ANR-11-IDEX-0004-02; Germany -- Bundesministerium f\"ur Bildung und Forschung (BMBF); Deutsche Forschungsgemeinschaft (DFG); Finanzministerium Baden-W\"urttemberg; Helmholtz Alliance for Astroparticle Physics (HAP); Helmholtz-Gemeinschaft Deutscher Forschungszentren (HGF); Ministerium f\"ur Innovation, Wissenschaft und Forschung des Landes Nordrhein-Westfalen; Ministerium f\"ur Wissenschaft, Forschung und Kunst des Landes Baden-W\"urttemberg; Italy -- Istituto Nazionale di Fisica Nucleare (INFN); Istituto Nazionale di Astrofisica (INAF); Ministero dell'Istruzione, dell'Universit\'a e della Ricerca (MIUR); CETEMPS Center of Excellence; Ministero degli Affari Esteri (MAE); Mexico -- Consejo Nacional de Ciencia y Tecnolog\'\i{}a (CONACYT) No.\ 167733; Universidad Nacional Aut\'onoma de M\'exico (UNAM); PAPIIT DGAPA-UNAM; The Netherlands -- Ministerie van Onderwijs, Cultuur en Wetenschap; Nederlandse Organisatie voor Wetenschappelijk Onderzoek (NWO); Stichting voor Fundamenteel Onderzoek der Materie (FOM); Poland -- National Centre for Research and Development, Grants No.\ ERA-NET-ASPERA/01/11 and No.\ ERA-NET-ASPERA/02/11; National Science Centre, Grants No.\ 2013/08/M/ST9/00322, No.\ 2013/08/M/ST9/00728 and No.\ HARMONIA 5--2013/10/M/ST9/00062, UMO-2016/22/M/ST9/00198; Portugal -- Portuguese national funds and FEDER funds within Programa Operacional Factores de Competitividade through Funda\c{c}\~ao para a Ci\^encia e a Tecnologia (COMPETE); Romania -- Romanian Authority for Scientific Research ANCS; CNDI-UEFISCDI partnership projects Grants No.\ 20/2012 and No.194/2012 and PN 16 42 01 02; Slovenia -- Slovenian Research Agency; Spain -- Comunidad de Madrid; Fondo Europeo de Desarrollo Regional (FEDER) funds; Ministerio de Econom\'\i{}a y Competitividad; Xunta de Galicia; European Community 7th Framework Program Grant No.\ FP7-PEOPLE-2012-IEF-328826; USA -- Department of Energy, Contracts No.\ DE-AC02-07CH11359, No.\ DE-FR02-04ER41300, No.\ DE-FG02-99ER41107 and No.\ DE-SC0011689; National Science Foundation, Grant No.\ 0450696; The Grainger Foundation; Marie Curie-IRSES/EPLANET; European Particle Physics Latin American Network; European Union 7th Framework Program, Grant No.\ PIRSES-2009-GA-246806; European Union's Horizon 2020 research and innovation programme (Grant No.\ 646623); and UNESCO.
\end{sloppypar}

\section*{Appendix A: Data Tables}

\setcounter{table}{0}
\renewcommand{\thetable}{A\arabic{table}}
\begin{table}[ht]
\caption{Values of $\langle$$\Delta_s$$\rangle$ for the 750 m array.
The fourth column shows the statistical uncertainty. For all
measurements the systematic uncertainty amounts to 0.07.}
\centering
\begin{tabular}{cccc} \hline\hline
Log (\textit{E}/eV) range & $\langle$Log (\textit{E}/eV) $\rangle$ &
$\langle$$\Delta_s$$\rangle$ &
$\sigma$\textsubscript{stat}($\langle$$\Delta_s$$\rangle$)\\
\hline
{[}17.5,17.6) & 17.55 & -0.157 & 0.009\\
{[}17.6,17.7) & 17.65 & -0.064 & 0.009\\
{[}17.7,17.8) & 17.75 & 0.004 & 0.008\\
{[}17.8,17.9) & 17.85 & 0.077 & 0.011\\
{[}17.9,18.0) & 17.95 & 0.170 & 0.014\\
{[}18.0,18.1) & 18.05 & 0.35 & 0.02\\
{[}18.1,18.2) & 18.15 & 0.41 & 0.03\\
{[}18.2,18.3) & 18.25 & 0.40 & 0.03\\
{[}18.3,18.4) & 18.35 & 0.54 & 0.03\\
{[}18.4,18.5) & 18.45 & 0.53 & 0.05\\
\hline\hline
\end{tabular}
\end{table}

\begin{table}
\caption{Values of $\langle$$\Delta_s$$\rangle$ for the 1500 m array.
The fourth column shows the statistical uncertainty. For all
measurements the systematic uncertainty amounts to 0.11.}
\centering
\begin{tabular}{cccc} \hline\hline
Log (\textit{E}/eV) range & $\langle$Log (\textit{E}/eV) $\rangle$ &
$\langle$$\Delta_s$$\rangle$ &
$\sigma$\textsubscript{stat}($\langle$$\Delta_s$$\rangle$)\\
\hline
{[}18.5,18.6) & 18.55 & -0.297 & 0.005\\
{[}18.6,18.7) & 18.65 & -0.242 & 0.006\\
{[}18.7,18.8) & 18.75 & -0.218 & 0.007\\
{[}18.8,18.9) & 18.85 & -0.163 & 0.009\\
{[}18.9,19.0) & 18.95 & -0.108 & 0.011\\
{[}19.0,19.1) & 19.05 & -0.056 & 0.012\\
{[}19.1,19.2) & 19.15 & 0.004 & 0.015\\
{[}19.2,19.3) & 19.25 & 0.077 & 0.020\\
{[}19.3,19.4) & 19.35 & 0.15 & 0.03\\
{[}19.4,19.5) & 19.45 & 0.11 & 0.03\\
{[}19.5,19.6) & 19.55 & 0.29 & 0.04\\
{[}19.6,19.7) & 19.64 & 0.20 & 0.04\\
{[}19.7,19.8) & 19.74 & 0.41 & 0.06\\
{[}19.8,$\infty$) & 19.88 & 0.60 & 0.06\\
\hline\hline
\end{tabular}
\end{table}

\begin{table*}
\caption{Values of $\langle$\textit{X}\textsubscript{max}$\rangle$ for the 750 m array.
The fourth column shows the statistical uncertainty. For all
measurements the systematic uncertainty amounts to 14 g
cm\textsuperscript{-2}.}
\centering
\begin{tabular}{cccc} \hline\hline
Log (\textit{E}/eV) range & $\langle$Log (\textit{E}/eV) $\rangle$ &
$\langle$\textit{X}\textsubscript{max}/ g cm\textsuperscript{-2}$\rangle$ &
$\sigma$\textsubscript{stat}($\langle$\textit{X}\textsubscript{max}$\rangle$)/g
cm\textsuperscript{-2}\\
\hline
{[}17.5,17.6) & 17.55 & 687.2 & 0.5\\
{[}17.6,17.7) & 17.65 & 695.6 & 0.6\\
{[}17.7,17.8) & 17.75 & 699.9 & 0.8\\
{[}17.8,17.9) & 17.85 & 707 & 1.0\\
{[}17.9,18.0) & 17.95 & 716 & 1.0\\
{[}18.0,18.1) & 18.05 & 733 & 2.0\\
{[}18.1,18.2) & 18.15 & 738 & 3.0\\
{[}18.2,18.3) & 18.25 & 745 & 3.0\\
{[}18.3,18.4) & 18.35 & 759 & 4.0\\
{[}18.4,18.5) & 18.45 & 754 & 5.0\\
\hline\hline
\end{tabular}
\end{table*}

\begin{table*}
\caption{Values of $\langle$\textit{X}\textsubscript{max}$\rangle$ for the 1500 m array.
The fourth column shows the statistical uncertainty. For all
measurements the systematic uncertainty amounts to 11 g
cm\textsuperscript{-2}.}
\centering
\begin{tabular}{cccc} \hline\hline
Log (\textit{E}/eV) range & $\langle$Log (\textit{E}/eV) $\rangle$ &
$\langle$\textit{X}\textsubscript{max}/ g cm\textsuperscript{-2}$\rangle$ &
$\sigma$\textsubscript{stat}($\langle$\textit{X}\textsubscript{max}$\rangle$) /g
cm\textsuperscript{-2}\\
\hline
{[}18.5,18.6) & 18.55 & 750.7 & 0.3\\
{[}18.6,18.7) & 18.65 & 755.2 & 0.3\\
{[}18.7,18.8) & 18.75 & 756.4 & 0.4\\
{[}18.8,18.9) & 18.85 & 759.8 & 0.6\\
{[}18.9,19.0) & 18.95 & 763.0 & 0.6\\
{[}19.0,19.1) & 19.05 & 766.5 & 0.7\\
{[}19.1,19.2) & 19.15 & 769.6 & 0.9\\
{[}19.2,19.3) & 19.25 & 775 & 1.0\\
{[}19.3,19.4) & 19.35 & 780 & 2.0\\
{[}19.4,19.5) & 19.45 & 779 & 2.0\\
{[}19.5,19.6) & 19.55 & 788 & 2.0\\
{[}19.6,19.7) & 19.64 & 785 & 2.0\\
{[}19.7,19.8) & 19.74 & 795 & 3.0\\
{[}19.8,$\infty$) & 19.88 & 807 & 3.0\\
\hline\hline
\end{tabular}
\end{table*}

\end{document}